\documentclass{statsoc}
\usepackage{amsmath,amsfonts,amssymb}
\usepackage{dsfont}
\usepackage{graphicx}
\usepackage{natbib}
\usepackage{algorithm}
\usepackage{algorithmic}
\usepackage[hyphens]{url}
\usepackage[a4paper]{geometry}
\usepackage{xcolor}

\title[Automated calibration for stability selection in penalised regression and graphical models]{Automated calibration for stability selection in penalised regression and graphical models}
  
\author[B Bodinier {\it et al.}]{Barbara Bodinier}
\address{MRC Centre for Environment and Health, Department of Epidemiology and Biostatistics, School of Public Health, Imperial College London, London, United Kingdom.}
\email{b.bodinier@imperial.ac.uk}

\author{Sarah Filippi}
\address{Department of Mathematics, Imperial College London, London, UK.}
\author{Therese Haugdahl Nøst}
\address{Systemsepidemiology, Department of Community Medicine, UiT The Arctic university of Norway, Troms\o, Norway.}
\author{Julien Chiquet}
\address{Universit\'e Paris-Saclay, AgroParisTech INRAE, UMR MIA, Paris, France.}
\author[B Bodinier {\textit et al.}]{Marc Chadeau-Hyam}
\address{MRC Centre for Environment and Health, Department of Epidemiology and Biostatistics, School of Public Health, Imperial College London, London, United Kingdom.}

\begin{document}

\begin{abstract}
Stability selection represents an attractive approach to identify sparse sets of features jointly associated with an outcome in high-dimensional contexts. We introduce an automated calibration procedure via maximisation of an in-house stability score and accommodating {\sl{a priori}}-known block structure (e.g. multi-OMIC) data. It applies to (LASSO) penalised regression and graphical models. Simulations show our approach outperforms non-stability-based and stability selection approaches using the original calibration. Application to multi-block graphical LASSO on real (epigenetic and transcriptomic) data from the Norwegian Women and Cancer study reveals a central/credible and novel cross-OMIC role of LRRN3 in the biological response to smoking. Proposed approaches were implemented in the R package sharp. 

\end{abstract}


\newpage

\section{Introduction}
\label{sec:intro}

% Smoking and OMICs
Tobacco smoking has long been established as a dangerous exposure causally linked to several severe chronic conditions. It has been estimated that one in five deaths in the United States was due to smoking \citep{CDCsmoking}. Nevertheless, the molecular mechanisms triggered and dysregulated by the exposure to tobacco smoking remain poorly understood. Over the past two decades, OMICs technologies have developed as valuable tools to explore molecular alterations due to external stressors or exposures \citep{Exposome}. Statistical analysis of OMICs data has enabled the identification of molecular markers of exposure at a single molecular level \citep{LondonSmoking, HuanSmoking} and are progressively moving towards the integration of data arising from different platforms \citep{Florence, ReviewMO}. There is an increasing need for efficient multivariate approaches accommodating high-dimensional and heterogeneous data typically exhibiting block-correlation structures. In particular, variable selection models can identify sparse sets of predictors and have proved useful for signal prioritisation in this context \citep{Chadeau2013, sPLSLymphoma}. Of these, the Least Absolute Shrinkage Selection Operator (LASSO) uses the $\ell_1$-penalisation of the coefficients to achieve variable selection \citep{lassoTibshirani}. Extensions of these penalised regression models have been proposed for the estimation of Gaussian graphical models \citep{graphicallassoMB, graphicallassoTibshirani}. By applying a $\ell_1$-penalisation to the precision matrix (as defined by the inverse of the covariance matrix), the graphical LASSO identifies non-zero entries of the partial correlation matrix. The evaluation (and subsequent selection) of pairwise relationships between molecular features in graphical models can guide biological interpretation of the results, under the assumption that statistical correlations reflect molecular interactions \citep{NetworkBiology, ValcarcelDN}. 

% Calibration of penalised models (state-of-the-art)
We focus in the present paper on the calibration of feature selection models, where feature denotes interchangeably a variable (in the context of regression) or an edge (graphical model). We illustrate our approach with regularised models, in which the model size (number of selected features) is controlled by the penalty parameter. The choice of parameter has strong implications on the generated results. Calibration procedures using cross-validation \citep{CVlasso, accuracycalib} or maximisation of information theory metrics, including the Bayesian (BIC) or Akaike (AIC) Information Criterion \citep{AIC, BIC, EBIC, GGMselect1} have been proposed. 

% Stability in penalised models
These models can be complemented by stability approaches to enhance reliability of the findings \citep{stabilityselectionMB, stabilityselectionSS, StARS}. In stability selection, the selection algorithm is combined with resampling techniques to identify the most stable signals. The model relies on the introduction of a second parameter: a threshold in selection proportion above which the corresponding feature is considered stable. A formula providing the upper-bound of the expected number of falsely selected features, or Per-Family Error Rate (PFER), as a function of the two parameters has been derived and is currently used to guide calibration \citep{stabilityselectionMB, stabilityselectionSS}. However, this calibration relies on the arbitrary choice of one of the two parameters, which can sometimes be difficult to justify.

% Our contribution (stats)
We introduce a score measuring the overall stability of the set of selected features, and use it to propose a new calibration strategy for stability selection. Our intuition is that all features would have the same probability of being selected in an unstable model. Our calibration procedure does not rely on the arbitrary choice of any parameter. Optionally, the problem can be constrained on the expected number of falsely selected variables and generate sparser results with error control.

% Challenge with multi-OMICs data
We also extend our calibration procedure to accommodate multiple blocks of data. This extension was motivated by the practical example on integration of data from different OMICs platforms. In this setting, block patterns arise, typically with higher (partial) correlations within a platform than between \citep{ChallengesMO}. We propose here an extension of stability selection combined with the graphical LASSO accommodating data with a known block structure. For this approach, each block is tuned using a block-specific pair of parameters (penalty and selection proportion threshold) \citep{LatentStructure}. 

% Analytical plan
We conduct an extensive simulation study to evaluate the performances of our calibrated stability selection models and compare them to state-of-the-art approaches. Our multi-OMICs stability-enhanced graphical models are applied to targeted methylation and gene expression data from an existing cohort. These datasets are integrated in order to characterise the molecular response to tobacco smoking at multiple molecular levels. The transcript of the LRRN3 gene, and its closest CpG site were found to play a central role in the generated graph. These two variables have the largest numbers of cross-OMICs edges and appear to be linking two largely uni-OMICs modules. LRRN3 methylation and gene expression therefore appear as pivotal molecular signals driving the biological response to tobacco smoking.

\section{Methods}
\label{sec:meth}

\subsection{Data overview}

We used DNA methylation and gene expression data in plasma samples from 251 women from the Norwegian Women and Cancer (NOWAC) cohort study \citep{NOWACIntegration}. Our study population includes 125 future cases (mean time-to-diagnosis of 4 years) and 126 healthy controls. The data was pre-processed as described elsewhere\citep{Florence}. DNA methylation at each CpG site are originally expressed as a proportion of methylated sequences across all copies ($\beta$-values) and was subsequent $\text{logit}_2$-transformed (M-values). The gene expression data was log-transformed. Features missing in more than 30\% of the samples were excluded, and the remaining data was imputed using the k-nearest neighbour. To remove technical confounding, the data was de-noised by extracting the residuals from linear mixed models with the OMIC feature as the outcome and modelling technical covariates (chip and position) as random intercepts \citep{NOWACIntegration}.

\subsection{Motivating research questions}
Our overarching research question is to identify the role of smoking-related CpG sites in lung carcingenesis and to better understand the molecular response to the exposure to tobacco smoke.

We therefore identified a subset of 160 CpG sites found differentially methylated in  never $vs.$ former smokers at a 0.05 Bonferroni corrected significance level in a large meta-analysis of 15,907 participants from 16 different cohorts \citep{LondonSmoking}. Similarly, we selected a set of 156 transcripts found differentially expressed in never $vs.$ current smokers from a meta-analysis including 10,233 participants from 6 cohorts \citep{HuanSmoking}. Of these, 159 CpG sites and 142 transcripts were assayed in our dataset.

Using a logistic-LASSO we first sought for a sparse subset of the (N=159) assayed smoking-related CpG sites that were jointly associated with the risk of future lung cancer. Second, to characterise the multi-OMICs response to exposure to tobacco smoking we estimated the conditional independence structure between smoking-related CpG sites (N=159) and transcripts (N=142) using the graphical LASSO.

To improve the reliability of our findings, both regularised regression and graphical models are used in a stability selection framework. These analyses raised two statistical challenges regarding the calibration of hyper-parameters in stability selection, and the integration of heterogeneous groups of variables in a graphical model. We detail below our approaches to accommodate these challenges.

\subsection{Variable selection with the LASSO}
\label{sec:lasso}

In LASSO regression, the $\ell_1$-penalisation is used to shrink the coefficients of variables that are not relevant in association with the outcome to zero \citep{lassoTibshirani}. Let $p$ denote the number of variables and $n$ the number of observations. Let $Y$ be the vector of outcomes of length $n$, and $X$ be matrix of predictors of size $(n \times p)$. The objective of the problem is to estimate the vector $\beta_{\lambda}$ containing the $p$ regression coefficients. The optimisation problem of the LASSO can be written:

\vspace{-15pt}
\begin{equation}
\min_{\beta_{\lambda}} ~\sum_{i=1}^n (y_i - \beta_{\lambda}^T x_i)^2 + \lambda \sum_{j=1}^p |\beta_{\lambda_j}|
\end{equation}

where $\lambda$ is a penalty parameter controlling the amount of shrinkage. \\

Penalised extensions of models including logistic, Poisson and Cox regressions have been proposed \citep{CoxLASSO}. In this paper, the use of our method is illustrated with LASSO-regularised linear regression. We use its implementation in the glmnet package in R (Gaussian family of models) \citep{CVlasso}. 

\subsection{Graphical model estimation with the graphical LASSO}
\label{sec:graphical_lasso}

A graph is characterised by a set of nodes (variables) and edges (pairwise links between them). As our data is cross-sectional, we focus here on undirected graphs without self-loops. As a result, the adjacency matrix encoding the network structure will be symmetric with zeros on the diagonal. \\

We assume that the data follows a multivariate Normal distribution: 

\vspace{-15pt}
\begin{equation}
X_i \sim N_p \big( \mu, \Sigma \big), ~~ i \in \{1, \dots, n\} 
\end{equation}
where $\mu$ is the mean vector and $\Sigma$ is the covariance matrix. \\

The conditional independence structure is encoded in the support of the precision matrix $\Omega = \Sigma^{-1}$. Various extensions of the LASSO have been proposed for the estimation of a sparse precision matrix \citep{graphicallassoMB, Banerjee}. We use here the graphical LASSO \citep{graphicallassoTibshirani} as implemented in the glassoFast package in R \citep{glasso, glassopackage, glassoFast}. For a given value of the penalty parameter $\lambda$, the optimisation problem can be written as: 

\vspace{-15pt}
\begin{equation}
\max_{\Omega}~ \text{log} ~\text{det} ~(\Omega) - \text{tr} ~(S \Omega) - \lambda || \Omega ||_{\ell_1}
\label{eq:glasso}
\end{equation}

where $S$ is the empirical covariance matrix and $ || \Omega ||_{\ell_1} =  \sum_{i\neq j} | \Omega_{ij} |$. \\

Alternatively, a penalty matrix $\Lambda$ can be used instead of the scalar $\lambda$ for more flexible penalisation: 

\vspace{-15pt}
\begin{equation}
\max_{\Omega}~ \text{log} ~\text{det} ~(\Omega) - \text{tr} ~(S \Omega) - || \Lambda \bullet \Omega ||_{\ell_1}
\label{eq:glassomat}
\end{equation}
where $\bullet$ denotes the element-wise matrix product. 

\subsection{Stability selection}
\label{sec:stability_selection}

Stability-enhanced procedures for feature selection proposed in the literature include stability selection \citep{stabilityselectionMB, stabilityselectionSS} and the Stability Approach to Regularization Selection (StARS) \citep{StARS}. Both use an existing selection algorithm and complement it with resampling techniques to estimate the probability of selection of each feature using its selection proportion over the resampling iterations. Stability selection ensures reliability of the findings through error control.

The feature selection algorithms we use are (a) the LASSO in a regression framework \citep{lassoTibshirani, CVlasso}, and (b) the graphical LASSO for the estimation of Gaussian graphical models \citep{graphicallassoMB, Banerjee, glassoFast} (see Supplementary Methods, section 1.1 and 1.2 for more details on the algorithms). The latter aims at the construction of a conditional independence graph. In a graph with $p$ nodes, for each pair of variables $X, Y$ and Gaussian vector $Z$ compiling the $(p-2)$ other variables, an edge is included if the conditional covariance $\mathrm{cov}(X, Y | Z)$ is different from zero (see Supplementary Methods, section 1.3 for more details on model calibration).

Under the assumption that the selection of feature $j$ is independent from the selection of any other feature $i \neq j$, the binary selection status of feature $j$ follows a Bernouilli distribution with parameter $p_{\lambda} (j)$, the selection probability of feature $j$. The stability selection model is then defined as the set $V_{\lambda, \pi}$ of features with selection probability above a threshold $\pi$:

\begin{equation}
V_{\lambda, \pi} = \{ j : p_{\lambda} (j) \geq \pi \}
\label{single}
\end{equation}

For each feature $j$, the selection probability is estimated as the selection proportion across models with penalty parameter $\lambda$ applied on $K$ subsamples of the data. 

The stability selection model has two parameters $(\lambda, \pi)$ that need to be calibrated. In the original paper, \citeauthor{stabilityselectionMB} use random subsamples of 50\% of the observations. They introduce $q_{\Lambda}$, the average number of features that are selected at least once by the underlying algorithm (e.g. LASSO) for a range of values $\lambda \in \Lambda$, across the $K$ subsamples. Under the assumptions of (a) exchangeability between selected features, and (b) that the selection algorithm is not performing worse than random guessing, they derived an upper-bound of the PFER, denoted by $\text{PFER}_{MB}$, as a function of the number of selected features $q_{\Lambda}$ and threshold in selection proportion $\pi$:
\[ 
\text{PFER}_{MB} (\Lambda, \pi) = \frac{1}{2 \pi - 1} \frac{q_{\Lambda}^2}{N} 
\]

With simultaneous selection in complementary pairs (CPSS), the selection proportions are obtained by counting the number of times the feature is selected in both the models fitted on a subsample of 50\% of the observations and its complementary subsample made of the remaining 50\% of observations \citep{stabilityselectionSS}. Using this subsampling procedure, the exchangeability assumption is not required for the upper-bound $\text{PFER}_{MB}$ to be valid. Under the assumption of unimodality of the distribution of selection proportions obtained with CPSS, \citeauthor{stabilityselectionSS} also proposed a stricter upper-bound on the expected number of variables with low selection probabilities, denoted here by $\text{PFER}_{SS}$: 
\begin{equation*}
\text{PFER}_{SS} (\Lambda, \pi) = 
\begin{cases}
\frac{1}{2 \times (2 \pi - 1 - 1 / K)} \frac{q_{\Lambda}^2}{N} ~~\text{if } \pi \leq 0.75 \\
\frac{4 \times (1 - \pi + 1 / K)}{1 + 2 / K} \frac{q_{\Lambda}^2}{N} ~~\text{otherwise.}
\end{cases}
\end{equation*}

For simplicity, we consider here point-wise control ($\Lambda$ reduces to a single value $\lambda$) with no effects on the validity of the formulas. Both approaches provide a relationship between $\lambda$ (via $q_{\lambda}$), $\pi$ and the upper-bound of the PFER such that if two of them are fixed, the third one can be calculated. The authors of both papers proposed to guide calibration based on the arbitrary choice of two of these three quantities. For example, the penalty parameter $\lambda$ can be calibrated for a combination of fixed values of the selection proportion $\pi$ and threshold in PFER. 

To avoid the arbitrary choice of the selection proportion $\pi$ or penalty $\lambda$, we introduce here a score measuring the overall stability of the model and use it to jointly calibrate these two parameters. We also consider the use of a user-defined threshold in PFER to limit the set of parameter values for $\lambda$ and $\pi$ to explore.

\subsection{Stability score}
\label{sec:stability_score}

Our calibration procedure aims at identifying the pair of hyper-parameters $(\lambda, \pi)$ that maximises model stability \citep{stability}. Let $H_{\lambda}(j) \in \{0, \dots, K\}$ denote the selection count of feature $j \in \{1, \dots, N\}$ calculated over the $K$ models fitted with parameter $\lambda$ over different subsamples. To quantify model stability, we first define three categories of features based on their selection counts. For a given penalty parameter $\lambda$ and threshold in selection proportion $\pi \in ] 0.5, 1 [$, each feature $j$ is either (a) stably selected if $H_{\lambda} (j) \geq K \pi$, (b) stably excluded if $H_{\lambda} (j) \leq K (1-\pi)$, or (c) unstably selected if $(1-\pi) K < H_{\lambda} (j) < K \pi$. Unstably selected features are those that are ambiguously selected across subsamples. The partitioning of the features into these three categories provides information about model stability, whereby a stable model would include a large numbers of stably selected and/or stably excluded features and a small number of unstably selected features.

We hypothesise that under the most unstable selection procedure, all features would have the same probability $\gamma_{\lambda} = q_{\lambda}/N$ of being selected, where $q_{\lambda} = \lfloor \frac{1}{K} \sum_{j = 1}^N H_{\lambda} (j) + \frac{1}{2} \rfloor$ is the average number of selected features across the $K$ models fitted with penalty $\lambda$ on the different subsamples of the data. Further assuming that the subsamples are independent, the selection count $H_{\lambda} (j)$ of feature $j \in \{1, \dots, N\}$ would then follow a binomial distribution:

\[
H_{\lambda} (j) \sim B \big( K, \gamma_{\lambda} \big)
\]

By considering the $N$ selection counts as independent observations, we can derive the likelihood of observing this classification under the hypothesis of instability, given $\lambda$ and $\pi$: 

\begin{align*}
L_{\lambda, \pi} = \prod_{j=1}^N \Big[ & \Big( 1 - F_{K, \gamma_{\lambda}} ( K \pi - 1 ) \Big)^{\mathds{1}_{ \{ H_{\lambda} (j) \geq K \pi \} }} \nonumber \\
\times & \Big( F_{K, \gamma_{\lambda}} ( K \pi - 1 ) - F_{K, \gamma_{\lambda}} ( K (1 - \pi)) \Big)^{\mathds{1}_{ \{ (1-\pi) K < H_{\lambda} (j) < K \pi \} }} \\ 
\times & F_{K, \gamma_{\lambda}} ( K ( 1 - \pi) )^{\mathds{1}_{ \{ H_{\lambda} (j) \leq K (1-\pi) \} }} \nonumber
 \Big]
\end{align*}

where $F_{K, \gamma_{\lambda}}$ is the cumulative probability function of the binomial distribution with parameters $K$ and $\gamma_{\lambda}$. 

Our stability score $S_{\lambda,\pi}$ is defined as the negative log-likelihood under the hypothesis of equi-probability of selection:
\[
S_{\lambda, \pi} = - \text{log} (L_{\lambda, \pi})
\]

The score $S_{\lambda, \pi}$ measures how unlikely a given model is to arise from the null hypothesis, for a given set of $\lambda$ and $\pi$. As such, the higher the score, the more stable the set of selected features. By construction, this formula is accounting for (a) the total number of features $N$, (b) the number of iterations $K$, (c) the density of selected sets by the original procedure via $\lambda$, and (d) the level of stringency as measured by threshold $\pi$. The calibration approach we develop aims at identifying sets of parameters $\lambda$ and $\pi$ maximising our score:
\begin{equation}
\max_{\lambda, \pi}~ S_{\lambda, \pi}
\label{unconstrained}
\end{equation}

Furthermore, this calibration technique can be extended to incorporate some error control via a constraint ensuring that the expected number of false positives is below an a priori fixed threshold in PFER $\eta$:
\begin{equation}
\max_{\lambda, \pi}~ S_{\lambda, \pi} \quad
\text{such that}~ U_{\lambda, \pi} \leq \eta , \text{where}
\label{constrained}
\end{equation}
$U_{\lambda, \pi}$ is the upper-bound used for error control in existing strategies (i.e. $\text{PFER}_{MB}$ or $\text{PFER}_{SS}$) \citep{stabilityselectionMB, stabilityselectionSS}. 

In the following sections, the use of Equation \eqref{unconstrained} is referred to as unconstrained calibration, and that of Equation \eqref{constrained} as constrained calibration.

\subsection{Multi-block graphical models}
\label{sec:multi_block}

The combination of heterogeneous groups of variables can create technically-induced patterns in the estimated (partial) correlation matrix, subsequently inducing bias in the generated graphical models. This can be observed, for example, when integrating the measured levels of features from different OMICs platforms. The between-platform (partial) correlations are overall weaker than within platforms (Supplementary Figure S1). This makes the detection of bipartite edges more difficult. This structure is known a priori and does not need to be inferred from the data. Indeed, the integration of data arising from $G$ homogeneous groups of variables generates $B=\frac{G\times(G+1)}{2}$ two-dimensional blocks in the (partial) correlation matrix where variables are ordered by group \citep{LatentStructure}. 

To tackle this scaling issue, we propose to use and calibrate block-specific pairs of parameters, $\lambda_b$ and $\pi_b$ controlling the level of sparsity in block $b$. Let $E_b, b \in \{1, \dots, B\}$ denote the sets of edges belonging to each of the blocks, such that: 
\[
\bigcup\limits_{b=1}^B E_b = \{1, \dots, N\}\] 

The stability selection model can be defined more generally as:
\begin{equation}
V_{\lambda_1, \dots, \lambda_B, \pi_1, \dots, \pi_B} = \bigcup\limits_{b=1}^B \{ j \in E_b : p_{\lambda_1, \dots, \lambda_B} (j) \geq \pi_b \} \text{, where}
\label{multiparameter}
\end{equation}

The probabilities $p_{\lambda_1, \dots, \lambda_B} (j), j \in \{1, \dots, N\}$ are estimated as selection proportions of the edges obtained from graphical LASSO models fitted on $K$ subsamples of the data with a block penalty matrix such that edge $j \in E_b$ is penalised with $\lambda_b$. 

Our stability score is then defined, by block, as:
\begin{align*}
S_{\lambda_1, \dots, \lambda_B, \pi_1, \dots, \pi_B} & = -\log \Big( \prod_{b=1}^B \prod_{j \in E_b} \Big[ \Big( 1 - F_{K, \gamma_{\lambda_1, \dots, \lambda_B}} ( K \pi_b - 1 ) \Big)^{\mathds{1}_{ \{ H^b_{\lambda_1, \dots, \lambda_B} (j) \geq K \pi_b \} }} \nonumber \\
& \times \Big( F_{K, \gamma_{\lambda_1, \dots, \lambda_B}} ( K \pi_b - 1 ) - F_{K, \gamma_{\lambda_1, \dots, \lambda_B}} ( K (1 - \pi_b)) \Big)^{\mathds{1}_{ \{ (1-\pi_b) K < H^b_{\lambda_1, \dots, \lambda_B} (j) < K \pi_b \} }} \\ 
& \times F_{K, \gamma_{\lambda_1, \dots, \lambda_B}} ( K ( 1 - \pi_b) )^{\mathds{1}_{ \{ H^b_{\lambda_1, \dots, \lambda_B} (j) \leq K (1-\pi_b) \} }} \Big] \Big) \nonumber
\end{align*}

% \begin{align*}
% S_{\lambda_1, \dots, \lambda_B, \pi_1, \dots, \pi_B} & = -\log \Big( \prod_{b=1}^B \prod_{j \in E_b} \Big[ \mathbb{P} \big( H^b_{\lambda_1, \dots, \lambda_B} (j) \geq K \pi_b \big)^{\mathds{1}_{ \{ H^b_{\lambda_1, \dots, \lambda_B} (j) \geq K \pi_b \} }} \nonumber \\
% & \times \mathbb{P} \big( (1-\pi_b) K < H^b_{\lambda_1, \dots, \lambda_B} (j) < K \pi_b \big)^{\mathds{1}_{ \{ (1-\pi_b) K < H^b_{\lambda_1, \dots, \lambda_B} (j) < K \pi_b \} }} \\ 
% & \times \mathbb{P} \big( H^b_{\lambda_1, \dots, \lambda_B} (j) \leq K (1-\pi_b) \big)^{\mathds{1}_{ \{ H^b_{\lambda_1, \dots, \lambda_B} (j) \leq K (1-\pi_b) \} }} \Big] \Big) \nonumber
% \end{align*}

Alternatively, we propose a block-wise decomposition, as described in Equation \eqref{multiblock}. To ensure that pairwise partial correlations in each block are estimated conditionally on all other $(p-2)$ nodes, we propose to estimate them from graphical LASSO models where the other blocks are weakly penalised (i.e. with small penalty $\lambda_0$). We introduce $p^b_{\lambda_b, \lambda_0} (j)$ and $H^b_{\lambda_b, \lambda_0} (j)$, the selection probability and count of edge $j \in E_b$ as obtained from graphical LASSO models fitted with a block penalty matrix such that edges $j \in E_b$ are penalised with $\lambda_b$ and edges $i \in E_{\ell}, \ell \neq b$ are penalised with $\lambda_0$. We define the multi-block stability selection graphical model as the union of the sets of block-specific stable edges:

\begin{equation}
V_{\lambda_1, \dots, \lambda_B, \lambda_0, \pi_1, \dots, \pi_B} = \bigcup\limits_{b=1}^B \{ j \in E_b : p^b_{\lambda_b, \lambda_0} (j) \geq \pi_b \}
\label{multiblock}
\end{equation}

The pair of parameters is calibrated for each of the blocks separately using a block-specific stability score defined by:
\begin{align*}
S^b_{\lambda_b, \lambda_0, \pi_b} & = -\log \Big( \prod_{j \in E_b} \Big[ \Big( 1 - F_{K, \gamma_{\lambda_b, \lambda_0}} ( K \pi_b - 1 ) \Big)^{\mathds{1}_{ \{ H^b_{\lambda_b, \lambda_0} (j) \geq K \pi_b \} }} \nonumber \\
& \times \Big( F_{K, \gamma_{\lambda_b, \lambda_0}} ( K \pi_b - 1 ) - F_{K, \gamma_{\lambda_b, \lambda_0}} ( K (1 - \pi_b)) \Big)^{\mathds{1}_{ \{ (1-\pi_b) K < H^b_{\lambda_b, \lambda_0} (j) < K \pi_b \} }} \\ 
& \times F_{K, \gamma_{\lambda_b, \lambda_0}} ( K ( 1 - \pi_b) )^{\mathds{1}_{ \{ H^b_{\lambda_b, \lambda_0} (j) \leq K (1-\pi_b) \} }} \Big] \Big) \nonumber
\end{align*}

where $\gamma_{\lambda_b, \lambda_0}$ is calculated based on the selection counts in $H^b_{\lambda_b, \lambda_0}$. 

% \begin{align*}
% S^b_{\lambda_b, \lambda_0, \pi_b} & = -\log \Big( \prod_{j \in E_b} \Big[ \mathbb{P} \big( H^b_{\lambda_b, \lambda_0} (j) \geq K \pi_b \big)^{\mathds{1}_{ \{ H^b_{\lambda_b, \lambda_0} (j) \geq K \pi_b \} }} \nonumber \\
% & \times \mathbb{P} \big( (1-\pi_b) K < H^b_{\lambda_b, \lambda_0} (j) < K \pi_b \big)^{\mathds{1}_{ \{ (1-\pi_b) K < H^b_{\lambda_b, \lambda_0} (j) < K \pi_b \} }} \\ 
% & \times \mathbb{P} \big( H^b_{\lambda_b, \lambda_0} (j) \leq K (1-\pi_b) \big)^{\mathds{1}_{ \{ H^b_{\lambda_b, \lambda_0} (j) \leq K (1-\pi_b) \} }} \Big] \Big) \nonumber
% \end{align*}

The implication of these assumptions are evaluated by comparing the two approaches described in Equations \eqref{multiblock} and \eqref{multiparameter} in a simulation study.

\subsection{Implementation}

The stability selection procedure is applied for different values of $\lambda$ and $\pi$ and the stability score is computed for all visited pairs of parameters. The grid of $\lambda$ values is chosen so that the underlying selection algorithm visits a range of models from empty to dense (up to $50\%$ of edges selected by the graphical LASSO) \citep{CVlasso, StARSpulsar}. Values of the threshold $\pi$ vary between 0.6 and 0.9, as proposed previously \citep{stabilityselectionMB}.

\subsection{Simulation models}

In order to evaluate the performances of our approach and compare to other established calibration procedures, we simulated several datasets according to the models described below, which we implemented in the R package fake (version 1.3.0).

\subsubsection{Graphical models}
\label{sec:simul_graphical}

We build upon previously proposed models to simulate multivariate Normal data with an underlying graph structure \citep{huge}. Our contributions include (a) a procedure for the automated choice of the parameter ensuring that the generated correlation matrix has contrast, and (b) the simulation of block-structured data.

First, we simulate the binary adjacency matrix $\Theta$ of size $(p \times p)$ of a random graph with density $\nu$ using the Erd\"{o}s-R\'{e}nyi model \citep{ErdosRenyi} or a scale-free graph using the Barab\'{a}si-Albert preferential attachment algorithm \citep{BarabasiAlbert, huge}. To introduce a block structure in the generate data, the non-diagonal entries of the precision matrix $\Omega$ are simulated such that:
\begin{equation*}
\Omega_{ij} = 
\begin{cases}
0~~\text{if}~ \Theta_{ij} = 0 \\
\alpha_{ij} ~~\text{if}~\Theta_{ij} = 1 ~\text{and $i$ and $j$ belong to the same platform} \\
\alpha_{ij} v_{b}~~\text{if}~\Theta_{ij} = 1 ~\text{and $i$ and $j$ belong to different platforms.}
\end{cases} , ~~ i \neq j \text{ where}\nonumber
\end{equation*}
$\alpha_{ij} \sim \mathnormal{U}(\{-1, 1\})$ and $v_b \in [0, 1]$ is a user-defined parameter.

We ensure that the generated precision matrix is positive definite via diagonal dominance: 
\[
\Omega_{ii} = \sum_{j=1}^p | \Omega_{ij} | + u, \text{ where}
\]
$u>0$ is a parameter to be tuned. 

\noindent The data is simulated from the centered multivariate Normal distribution with covariance $\Omega^{-1}$. 

The simulation model is controlled by five parameters:
\begin{enumerate}
\item number of observations $n$,
\item number of nodes $p$,
\item density of the underlying graph $\nu \in [0, 1]$,
\item scaling factor $v_b \in [0, 1]$ controlling the level of heterogeneity between blocks,
\item constant $u>0$ ensuring positive definiteness.
\end{enumerate}

We propose to choose $u$ so that the generated correlation matrix has a high contrast, as defined by the number of unique truncated correlation coefficients with three digits (Supplementary Figure S2). The parameter $v_b \in [0, 1]$ is set to 1 (no block structure) for single-block simulations and chosen to generate data with a visible block structure for multi-block simulations ($v_b = 0.2$). These models generate realistic correlation matrices (Supplementary Figure S1).

\subsubsection{Linear regression}

For linear regression, the data simulation is done in two steps with (i) the simulation of $n$ observations for the $p$ predictors, and (ii) the simulation of the outcome for each of the $n$ observations, conditionally on the predictors data. The first step is done using the simulation model introduced in the previous section for graphical models. This allows for some flexibility over the (conditional) independence patterns between predictors. For the second step, we sample $\beta$-coefficients from a uniform distribution over $\{ -1, 1 \}$ (for homogeneous effects in absolute value) or over $\{ [-1, 0.5] \cup [0.5, 1] \}$ (to introduce variability in the strength of association with the outcome). The outcome $Y_i, i \in \{1, \dots, n\}$ is then sampled from a Normal distribution \citep{CVlasso}:

\[
Y_i | X_i = x_i \sim \mathcal{N} (x_i \beta, \sigma^2)
\]

The parameter $\sigma$ controls the proportion of variance in the outcome that can be explained by its predictors. The value of $\sigma$ is chosen to reach the expected proportion of explained variance $R^2$ used as simulation parameter:

\[
\sigma = \sqrt{\frac{1-R^2}{R^2}s^2}
\]
where $s^2$ is the variance of $X \beta$.

\subsubsection{Performance metrics}

Selection performances of the investigated models are measured in terms of precision $p$ and recall $r$: 

\[
p = \frac{TP}{TP + FP} ~~\text{and}~~ r = \frac{TP}{TP+FN} \text{, where}
\]
$TP$ and $FP$ are the numbers of true and false positives, respectively, and $FN$ is the number of false negatives.

The $F_1$-score quantifies the overall selection performance based on a single metric: 

\[
F_1=\frac{2 \times p \times r}{p+r}
\]

\section{Simulation study}
\label{sec:simul}

We use a simulation study to demonstrate the relevance of stability selection calibrated with our approach:
\begin{enumerate}
    \item in a linear regression context for the LASSO model,
    \item for graphical model using the graphical LASSO,
    \item for multi-block graphical models.
\end{enumerate}

From these we evaluate the relevance of our stability score for calibration purposes, and compare our score to a range of existing calibration approaches including information theory criteria, StARS, and stability selection models using the previously proposed error control for different values of the threshold in selection proportion $\pi$. As sensitivity analyses, we evaluate the performances of stability selection for graphical models using different resampling approaches, different numbers of iterations $K$, and compare the two proposed approaches for multi-block calibration.

\subsection{Simulation parameters}

All simulation parameters were chosen in an attempt to generate realistic data with many strong signals and some more difficult to detect (weaker partial correlation).

For graphical models, we used $p=100$ nodes with an underlying random graph structure of density $\nu=0.02$ (99 edges on average, as would be obtained in a scale-free graph with the same number of nodes). For multi-block graphical models, we considered two homogeneous groups of 50 nodes each. Reported distributions of selection metrics were computed over 1,000 simulated datasets.

Unless otherwise stated, stability selection models were applied on grids of 50 dataset-specific penalty parameter values and 31 values for the threshold in selection proportion between $0.6$ and $0.9$. The stability-enhanced models were based on $K=100$ (complementary) subsamples of $50\%$ of the observations.

\subsection{Applications to simulated data}

\begin{figure}[h!]
\centering
\makebox{\includegraphics[width=0.64\linewidth]{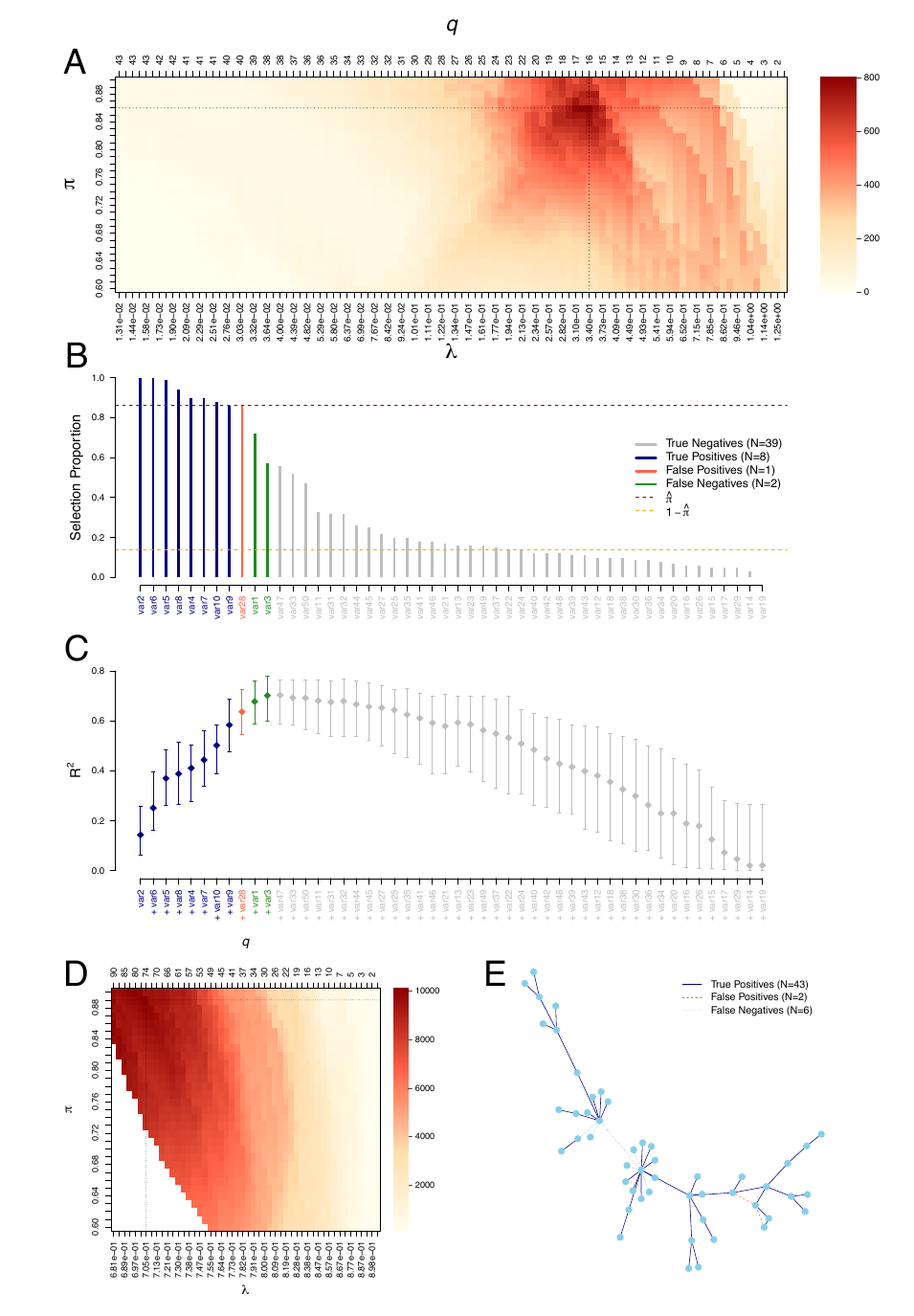}}
\caption{\textbf{Stability selection LASSO (A-C) and graphical LASSO (D-E) applied on simulated data.} Calibration plots (A, D) show the stability score (colour-coded) for different penalty parameters $\lambda$, or numbers of features selected $q$, and thresholds in selection proportion $\pi$. We show selection proportions (B) and a graph representation of the detected and missed edges (E). We report the median, $5^{th}$ and $95^{th}$ quantiles of the $R^2$ obtained for $100$ unpenalised regression models sequentially adding the predictors in order of decreasing selection proportions (C). These models are trained on $50\%$ of the data and performances are evaluated on the remaining observations. True Positives (dark blue), False Positives (red dashed line) and False Negatives (green dotted line) are highlighted (B, C, E). Calibration of the stability selection graphical LASSO ensures that the expected number of False Positives (PFER) is below $20$ (D). The two datasets are simulated for $p = 50$ variables and $n = 100$ observations. For the regression model, $10$ variables contribute to the definition of the outcome with effect sizes in $\{ [-1, -0.5] \cup [0.5, 1] \}$ and an expected proportion of explained variance of $70\%$. For the graphical model, the simulated graph is scale-free.}
\label{fig:illustration_lasso}
\end{figure}

Our stability selection approach is first applied to the LASSO for the selection of variables jointly associated with a continuous outcome in simulated data (Figure \ref{fig:illustration_lasso}). 

The penalty parameter $\lambda$ and threshold in selection proportion $\pi$ are jointly calibrated to maximise the stability score (Figure \ref{fig:illustration_lasso}-A). Stably selected variables are then identified as those with selection proportions greater than the calibrated parameter $\hat{\pi}=0.86$ (dark red line) in LASSO models fitted on 50\% of the data with calibrated penalty parameter $\hat{\lambda}=0.34$ (Figure \ref{fig:illustration_lasso}-B). The resulting set of stably selected variables includes 8 of the 10 'true' variables used to simulate the outcome and 1 'wrongly selected' variables we did not use in our simulation. 

We observe a marginal increase in prediction performances across unpenalised models sequentially adding the 9 stably selected predictors by order of decreasing selection proportions (Figure \ref{fig:illustration_lasso}-C). Further including the two False Negatives generates a limited increase in $R^2$, and so does the inclusion of any subsequent variable. This suggests that our stability selection model captures most of the explanatory information and was therefore well calibrated.

To limit the number of 'wrongly selected' features, we can restrict the values of $\lambda$ and $\pi$ visited so they ensure a control of the PFER (Supplementary Figure S3). In that constrained optimisation, the values of $\lambda$ and $\pi$ yielding a PFER exceeding the specified threshold are discarded and corresponding models are not evaluated (Supplementary Figure S3-A). The maximum stability score can be obtained for different pairs $(\lambda, \pi)$ depending on the constraint, but our simulation shows that differences in the maximal stability score (Supplementary Figure S3-B) and resulting selected variables are small (Supplementary Figure S3-C) if the constraint is not too stringent. 

Our stability score is also used to calibrate the graphical LASSO for the estimation of a conditional independence graph, while controlling the expected number of falsely selected edges below 20 (Figure \ref{fig:illustration_lasso}-C). The calibrated graph (Figure \ref{fig:illustration_lasso}-D) included 56 (47 rightly, in plain dark blue and 9 wrongly, in dashed red lines) stably selected edges (i.e. with selection proportions $\geq \hat{\pi}=0.90$), based on graphical LASSO models fitted on 50\% of the data with penalty parameter $\lambda=0.52$. The 9 wrongly selected edges tend to be between nodes that are otherwise connected in this example (marginal links). The 2 missed edges are connected to the central hub and thus correspond to smaller partial correlations, more difficult to detect.

\subsection{Evaluation of model performance and comparison with existing approaches}

\begin{figure}[h!]
\centering
\includegraphics[width=\linewidth]{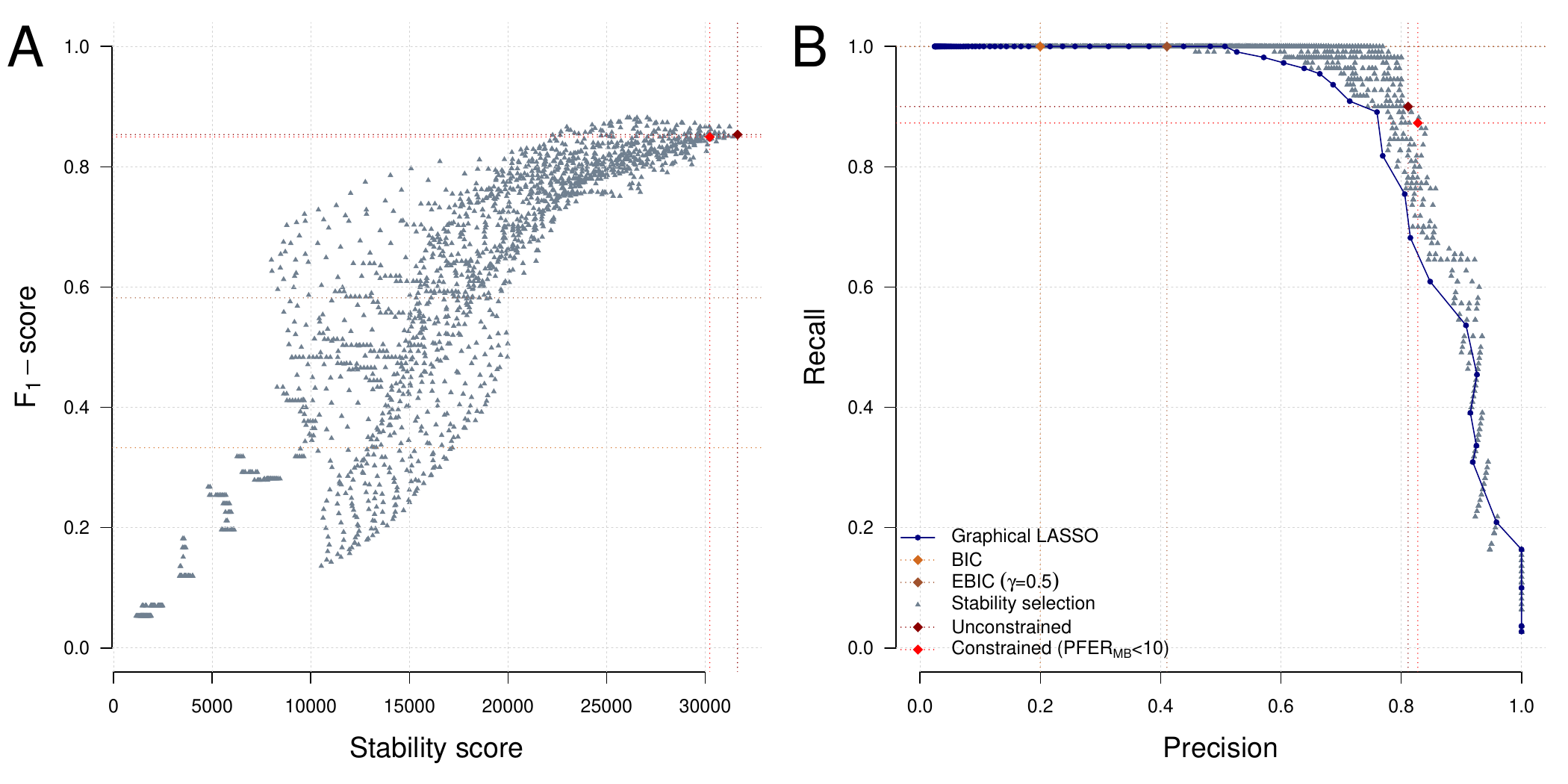} \\
\caption{\textbf{Selection performance in stability selection and relevance of the stability score for calibration.} The graphical LASSO and stability selection are applied on simulated data with $n = 200$ observations for $p = 100$ variables where the conditional independence structure is that of a random network with $\nu = 0.02$. The $F_1$-score of stability selection models fitted with a range of $\lambda$ and $\pi$ values is represented as a function of the stability score (A). Calibrated stability selection models using the unconstrained (dark red) and constrained (red) approaches are highlighted. The precision and recall of visited stability selection models (grey) and corresponding graphical LASSO models (dark blue) are reported (B). The calibrated models using the BIC (beige) or EBIC (brown) are also showed (B).}
\label{Fig_perf}
\end{figure}

Our simulations show that models with higher stability score yield higher selection performances (as measured by the $F_1$-score), making it a relevant metric for calibration (Figure \ref{Fig_perf}-A). We also find that irrespective of the value of $\lambda$ and $\pi$, stability selection models outperform the original implementation of the graphical LASSO (Figure \ref{Fig_perf}-B). Graphical LASSO calibrated using the BIC or EBIC (see Supplementary Methods, section 1.3) generate dense graphs resulting in perfect recall and poor precision values (0.20 and 0.41). Our stability score instead yield sparser models, resulting in slightly lower recall values (0.90) which did not include many irrelevant edges, as captured by the far better precision value (0.81). Our simulation also shows that the constraint controlling the PFER further improves the precision (0.83) through the generation of a sparser model. 

Our calibrated stability selection graphical LASSO models are compared with state-of-the-art graphical model estimation approaches on 1,000 simulated datasets in low, intermediate and high-dimension (Figure 3, Supplementary Table S1). Non stability-enhanced graphical LASSO models, calibrated using information theory criteria, are generally the worst performing models (median $F_1$-score $<0.6$ across dimensionality settings). StARS models, applied with the same number of subsampling iterations and using default values for other parameters, have the highest median numbers of True Positives. However, they include more False Positives than stability selection models, making it less competitive in terms of $F_1$-score (best performing in high-dimension with a median $F_1$-score of 0.66). For stability selection models calibrated using error control (MB \citep{stabilityselectionMB}, SS \citep{stabilityselectionSS}), the optimal choice of $\pi$ seems to depend on many parameters including the dimensionality and structure of the graph (Supplementary Figure S4). By jointly calibrating the two parameters, our models show generally better performances compared to models calibrated solely using error control on these simulations (median $F_1$-score ranging from 0.69 to 0.72 using $\text{PFER}_{SS}<20$ only in high dimension, compared to 0.74 using constrained calibration maximising the stability score). Results were consistent when using different thresholds in PFER (Supplementary Figure S5). For LASSO models, we observe a steep increase in precision with all stability selection models compared to models calibrated by cross-validation (Supplementary Figure S6). Unconstrained calibration using our stability score yielded the highest $F_1$-scores in the presence of independent or correlated predictors. Computation times of the reported stability selection models are comparable and acceptable in practice (less than 3 minutes in these settings) but rapidly increase with the number of nodes for graphical models, reaching 8 hours for 500 observations and 1,000 nodes (Supplementary Table S2).

\begin{figure}[h!]
\centering
\includegraphics[width=\linewidth]{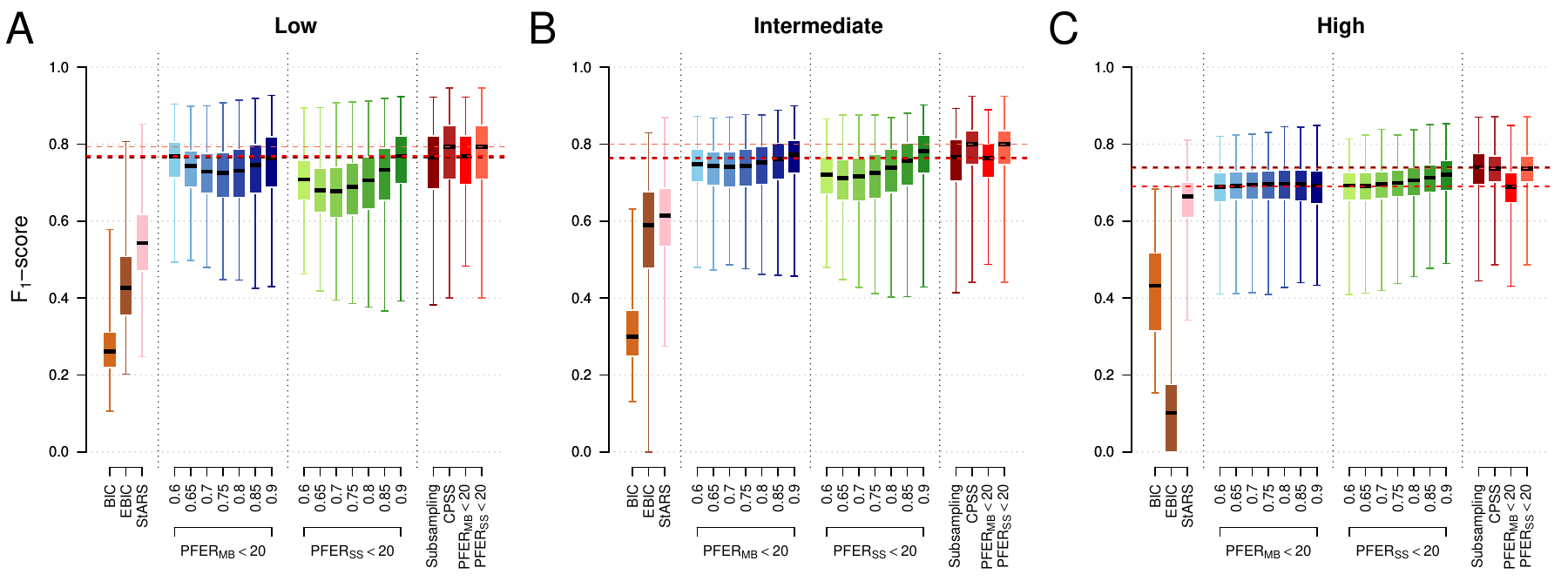} \\
\caption{\textbf{Selection performances of state-of-the-art approaches and proposed calibrated stability selection graphical LASSO models.} We show the median, quartiles, minimum and maximum $F_1$-score of graphical LASSO models calibrated using the BIC, EBIC, StARS, and stability selection graphical LASSO models calibrated via error control (MB in blue, SS in green) or using the proposed stability score (red). Models are applied on 1,000 simulated datasets with $p=100$ variables following a multivariate Normal distribution corresponding to a random graph structure ($\nu=0.02$). Performances are estimated in low ($n=2p=200$, A), intermediate ($n=p=100$, B), and high ($n=p/2=50$, C) dimensions.}
\end{figure}

\subsection{Sensitivity to the choice of resampling procedure}

Stability selection can be implemented with different numbers of iterations $K$ and resampling techniques (subsampling, bootstrap or CPSS approaches, and subsample size). We show in a simulation study with $p=100$ nodes that (a) the effect of the number of iterations $K$ reaches a plateau after $50$ of iterations, and (b) that the best performances were obtained for bootstrap samples or subsamples of 50\% of the observations (Supplementary Figures S7 and S8).

\subsection{Multi-block extension for graphical models}

Our single and multi-block calibration procedures are applied on simulated datasets with a block structure in different dimensionality settings. Block specific selection performances of both approaches can be visualised in precision-recall plots (Figure 4, Supplementary Table S3). Irrespective of the dimensionality, accounting for the block structure as proposed in Equation \eqref{multiblock} with $\lambda_0 = 0.1$ generates an increase in selection performance in both within and between blocks (up to 7\% in overall median $F_1$-score in low dimension). This gain in performance comes at the price of an increased computation time (from 2 to 6 minutes in low dimension). 

Additionally, we show in Supplementary Table S4 that the choice of $\lambda_0$ has limited effects on the selection performances, as long as it is relatively small ($\lambda_0 \leq 0.1$). We choose $\lambda_0 = 0.1$ for a good balance between performance and computation time. We also show that the use of Equation \eqref{multiblock} gives better selection performances than that of Equation \eqref{multiparameter} (median $\text{F}_1$-score $\geq 0.71$ compared to $0.57$). In particular, it drastically reduces the numbers of False Positives in the off-diagonal block.

\begin{figure}[h!]
\centering
\includegraphics[width=\linewidth]{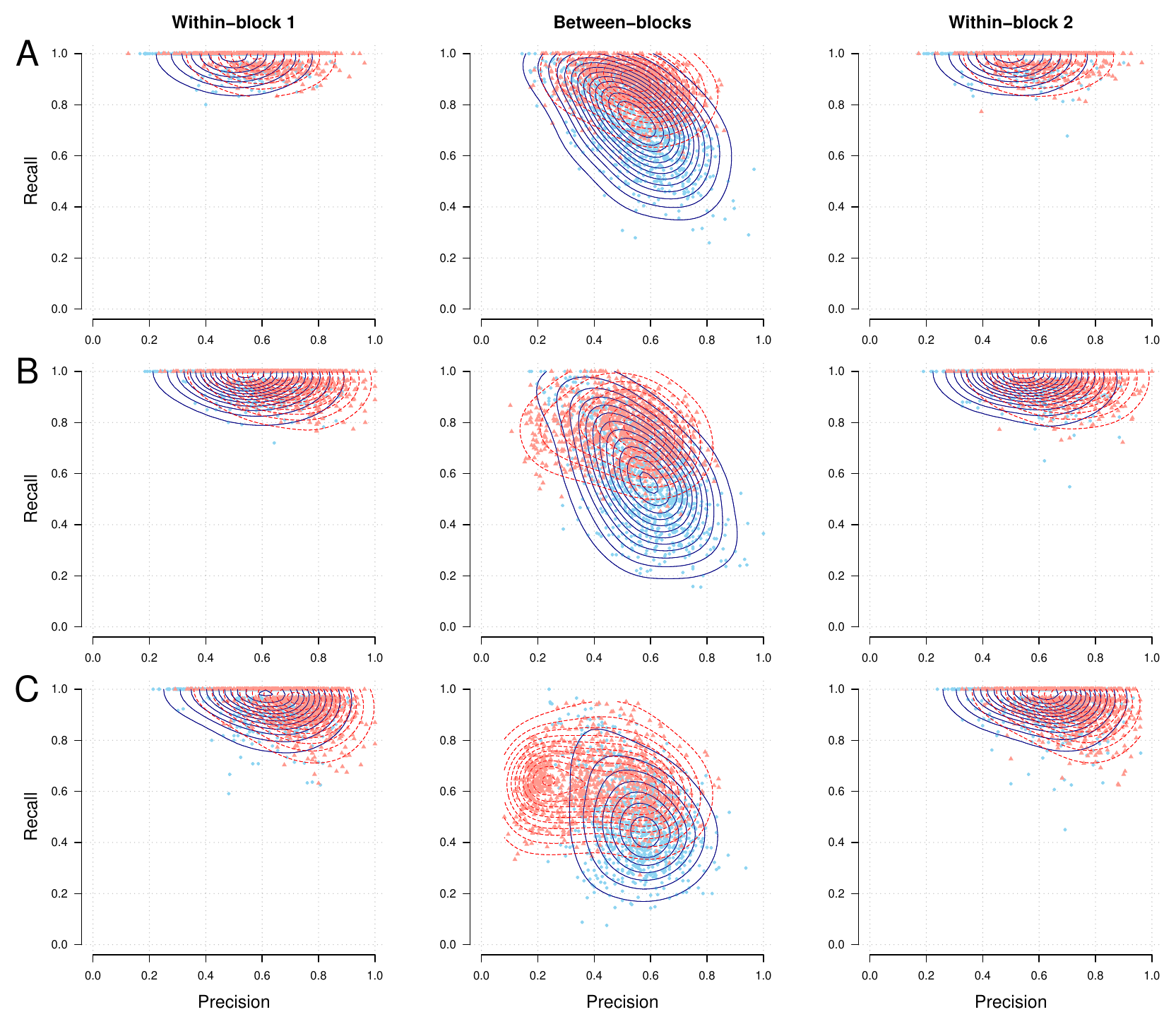} \\
\caption{\textbf{Precision-recall showing single and multi-block stability selection graphical models applied on simulated data with a block structure.} Models are applied on 1,000 simulated datasets (points) with $p=100$ variables following a multivariate Normal distribution corresponding to a random graph ($\nu=0.02$) and with known block structure ($50$ variables per group, using $v_b=0.2$). The contour lines indicate estimated 2-dimensional density distributions. Performances are evaluated in low (A, $n=2p=200$), intermediate (B, $n=p=100$), and high (C, $n=p/2=50$) dimensions.}
\end{figure}

\section{Application: molecular signature of smoking}
\label{sec:appli}

\subsection{Epigenetic markers of lung cancer}

To identify smoking-related markers that contribute to the risk of developing lung cancer, we use stability selection logistic-LASSO with the 159 CpG sites as predictors and the future lung cancer status as outcome (Figure 5-A, B). The calibrated model includes 21 CpG sites with selection proportions above 0.66. The unpenalised logistic models with stably selected predictors reach a median AUC of 0.69, which is close to that of pack years (median AUC of 0.74) and implies that these 21 CpG sites capture most of the information on smoking history relevant to lung cancer prediction. The limited increase in AUC beyond the calibrated number of predictors suggests that the stability selection model achieves a good balance between sparsity and prediction performance.

\subsection{Multi-OMICs graph}

We first estimate the conditional independence structure between smoking-related CpG sites with single-block stability selection (Supplementary Figure S9). A total of 320 edges involving 100 of the 159 CpG sites are obtained. Most CpG sites are in the same connected component, but we also observe 6 small modules made of 2 or 3 nodes. 

In order to get a more comprehensive understanding of the biological response to smoking we integrate methylation data, known to reflect long-term exposure to tobacco smoking, and gene expression, which is functionally well characterised, and seek for correlation patterns across these smoking-related signals via the estimation of a multi-OMICs graph.

\begin{figure}[h!]
\centering
\includegraphics[width=0.75\linewidth]{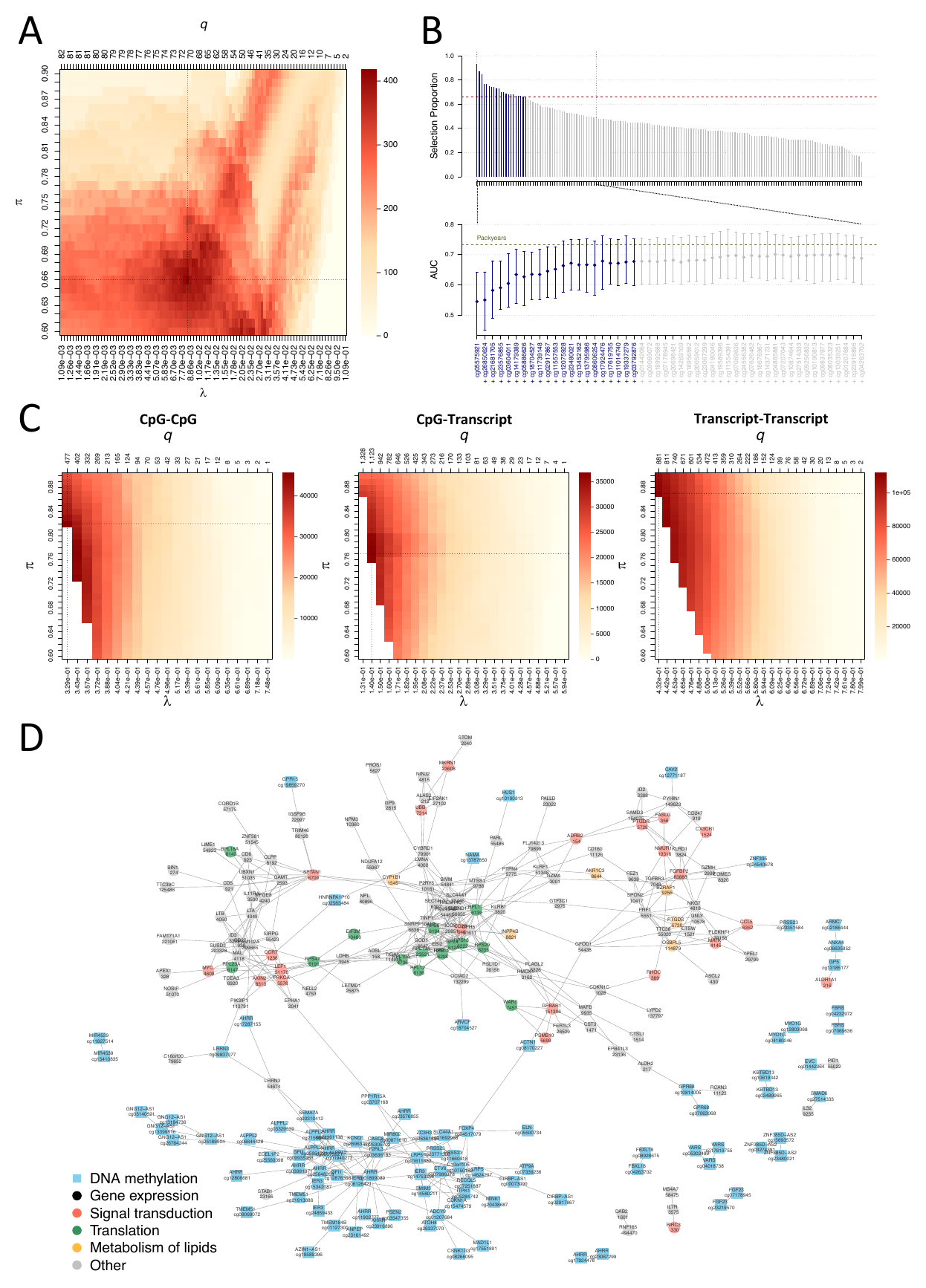} \\
\caption{\textbf{Stability selection on real DNA methylation and gene expression data.} The stability selection logistic-LASSO with the future lung cancer status as outcome and epigenetic markers of smoking as predictors is calibrated by maximising the stability score (A). The selection proportions in the calibrated model and explanatory performances of unpenalised logistic models where the predictors are sequentially added by decreasing selection proportion are showed (B). The three blocks of a multi-OMICs graphical model integrating DNA methylation and gene expression markers of tobacco smoking are calibrated separately using models where the other blocks are weakly penalised ($\lambda_0$=0.1), while ensuring that $\text{PFER}_{MB}<150$ overall (C). The stability selection model includes edges that are stably selected in each block (D).}
\end{figure}

We accommodate the heterogeneous data structure (Supplementary Figure S10) by calibrating three pairs of block-specific parameters $(\lambda, \pi)$ using our multi-block strategy (Figure 5-A). We found a total of 601 edges, including 150 in the within-methylation block, 425 in the within-gene expression block, and 26 cross-OMICs edges (Figure 5-B). The detected links reflect potential participation to common regulatory processes of both transcripts and CpG sites. As our analysis was limited to smoking-related markers, connected nodes can be hypothesised to jointly contribute to the biological response to tobacco smoking.

For comparison, we estimate the graphical LASSO model calibrated using the BIC on the same data (Supplementary Figure S11). Of the 601 edges included in the stability selection graph, 583 were also in the BIC-calibrated graph. The BIC-calibrated graph is more dense (N=6,744 edges), which makes it difficult to interpret. As this procedure does not account for the block structure in the data, two modules corresponding to the two platforms are clearly visible.

DNA methylation nodes are annotated with the symbol of its closest gene on the genome \citep{LondonSmoking}. Most sets of CpG sites annotated with the same gene symbol are interconnected in the graph (e.g. AHRR, GNG12-AS1, and ALPPL2 on chromosomes 5, 3, and 2, respectively). The data includes a CpG site and a transcript with the same annotation for two genes, but only found a cross-OMIC link for LRRN3 \citep{Florence}. The LRRN3 transcript, which is linked to 4 CpG sites including AHRR, ALPPL2 and a CpG site annotated as LRRN3 (cg09837977), has a central position among methylation markers (Figure 5-B). 

Strong correlations involving features that are closely located on the genome, or cis-effects, have been reported previously \citep{DMR}. Our approach also detects cross-chromosome edges (Supplementary Figure S12), suggesting that complex long-range mechanisms could be at stake \citep{MethylationMechanisms}. 

We incorporate functional information in the visualisation using Reactome pathways (Figure 5-B) \citep{WGCNA, Reactome}. As previously reported, the immune system and in signal transduction (red) pathways were largely represented in the targeted set \citep{HuanSmoking, NOWACIntegration}. Interestingly, the group of interconnected nodes around RPL4 (green) was involved in a range of pathways including the cellular response to stress, translation, and developmental biology. Similarly, the transcripts involved in the metabolism of lipids (yellow) are closely related in the graph. Altogether these results confirm the functional proximity of the nearby variables from our graph, hence lending biological plausibility of its topology.

\section{Discussion}
\label{sec:discussion}

The stability selection models and proposed calibration procedure have been implemented in the R package sharp (version 1.2.1), available on CRAN. The selection performances of our variable selection and (multi-block) graphical models were evaluated in a simulation study. We showed that stability selection models yield higher $\text{F}_1$-score, to the cost of a (limited) increase in computation time. The computational efficiency of the proposed approaches can easily be improved using warm start and parallelisation, both readily implemented in the R package sharp. We also demonstrated that the proposed calibration procedure is generally identifying the optimal threshold in selection proportion which leads to overall equivalent or better performances than previously proposed approaches based solely on error control. Our multi-block extension was successful in removing some of the technical bias through a more flexible modelling, but generated a ten-fold increase in computation time compared to single-block models on these simulations.

The proposed approaches also generated promising results on real OMICs data \citep{Dusan}. The development of stability-enhanced models accommodating data with a known block structure we proposed was triggered by the multi-OMICs application for the characterisation of the molecular signature of smoking. Their application to methylation and gene expression data gave further insights on the long-range correlations previously reported \citep{Florence}, and revealed a credible pivotal cross-OMICs role of the LRRN3 transcript \citep{HuanSmoking}. Annotation of the networks using biological information from the Reactome database identifies modules mostly composed of nodes belonging to the same pathways, suggesting that statistical correlations can reflect functional role in shared biological pathways.

The stability selection approach and calibration procedure introduced here could also be used in combination with other variable selection algorithms, including penalised unsupervised models that cannot rely on the minimisation of an error term in cross-validation \citep{sPCA}, or extensions modelling longitudinal \citep{TimeCourse} or count data \citep{PLNnetwork}. The method and its implementation in the R package sharp comes with some level of flexibility and user-controlled choices. Depending on the application and its requirements, the models can be tailored to generate more or less conservative results using (a) the threshold in PFER controlling the sparsity of the selected sets, and (b) considering features with intermediate selection proportions (between $1-\pi$ and $\pi$). The calculation of our stability score can alternatively be based on two categories including (a) stably selected features with $H_{\lambda} (j) \geq K \pi$, and (b) non-stably selected features with $H_{\lambda} (j) < K \pi$. As this definition would ignore stably excluded features, which also contribute to the overall model stability, it may hamper selection performances.

Nevertheless, the results of stability selection models should always be interpreted with care. Our simulation studies indicate that, even when the assumptions of the model are verified (including the multivariate Normal distribution), the estimations of the graphical models are not perfect and need to be interpreted with care. In particular, some of the edges selected may correspond to marginal relationships (and not true conditional links). On the other hand, the absence of an edge does not necessarily indicate that there is no conditional association between the two nodes (especially for cross-group edges, for which the signal is diluted). Reassuringly, the overall topology of the graph seems relevant, as observed when applied on data with a scale-free graphical structure.

As with all penalised approaches, the stability selection models we propose rely on a sparsity assumption. In regression, this assumption implies that some of the predictors do not contribute to the prediction of the outcome or provide information that is redundant with that from other predictors. As the stability score $S_{\lambda, \pi}$ we propose is equal to zero the stability selection model is empty (no stably selected features) or saturated (all features are stably selected), our calibration procedure is only informative for models where the number of stably selected features is between $1$ and $(N-1)$. The validity of this sparsity assumption could be investigated post-hoc using unpenalised models sequentially adding the selected features in decreasing order of selection proportion.

The calculation of the stability score relies on the assumption that the feature selection counts are independent. The link between correlation across features and correlation of their selection counts is not obvious and would warrant further investigation. However, selection and prediction performances of our calibrated stability selection LASSO models do not seem to be affected by the presence of correlated predictors

While stability selection LASSO has been successfully applied on high dimensional data with almost 450,000 predictors \citep{Dusan}, the stability selection graphical LASSO has limited scalability. The complexity of graphical models is rapidly increasing with the number of nodes, and despite recent faster implementations of the graphical models \citep{glassoFast}, computation times remain high with more than a few hundreds of nodes. Beyond their computational burden, large graphical models can become very dense and more efficient ways of visualising and summarising the results will be needed. Alternatively, as structures of redundant interconnected nodes (cliques) can be observed, summarising these in super-nodes could be valuable. This could be achieved using clustering or dimensionality reduction approaches, or by incorporating a priori biological knowledge in the model.

\section{Data Availability Statement}

Data sharing is not applicable to this article as no new data were created or analysed in this study. All codes and simulated datasets are available on \url{https://github.com/barbarabodinier/stability_selection}. The R packages sharp and fake are available on the Comprehensive R Archive Network (CRAN).

\section{Acknowledgements}

We are very grateful to Prof St\'ephane Robin for his insightful comments on the models and their applications. This work was supported by the Cancer Research UK Population Research Committee “Mechanomics” project grant (grant no. 22184 to M. Chadeau-Hyam). B. Bodinier received a PhD Studentship from the MRC Centre for Environment and Health. M. Chadeau-Hyam and T. Haugdahl Nøst acknowledge support from the Research Council of Norway (Id-Lung project FRIPRO 262111). J. Chiquet acknowledges support from ANR-18-CE45-0023 Statistics and Machine Learning for Single Cell Genomics (SingleStatOmics). M. Chadeau-Hyam acknowledges support from the H2020-EXPANSE project (Horizon 2020 grant no. 874627) and H2020-Longitools project (Horizon 2020 grant no. 874739).

\bibliographystyle{chicago}

\bibliography{biblio}

\begin{thebibliography}{}

\bibitem[\protect\citeauthoryear{Akaike}{Akaike}{1998}]{AIC}
Akaike, H. (1998).
\newblock {\em Information Theory and an Extension of the Maximum Likelihood
  Principle}, pp.\  199--213.
\newblock New York, NY: Springer New York.

\bibitem[\protect\citeauthoryear{Albert and Barab\'asi}{Albert and
  Barab\'asi}{2002}]{BarabasiAlbert}
Albert, R. and A.-L. Barab\'asi (2002, Jan).
\newblock Statistical mechanics of complex networks.
\newblock {\em Rev. Mod. Phys.\/}~{\em 74}, 47--97.

\bibitem[\protect\citeauthoryear{Ambroise, Chiquet, and Matias}{Ambroise
  et~al.}{2009}]{LatentStructure}
Ambroise, C., J.~Chiquet, and C.~Matias (2009).
\newblock Inferring sparse gaussian graphical models with latent structure.
\newblock {\em Electron. J. Statist.\/}~{\em 3}, 205--238.

\bibitem[\protect\citeauthoryear{Banerjee, El~Ghaoui, and
  d’Aspremont}{Banerjee et~al.}{2008}]{Banerjee}
Banerjee, O., L.~El~Ghaoui, and A.~d’Aspremont (2008, June).
\newblock Model selection through sparse maximum likelihood estimation for
  multivariate gaussian or binary data.
\newblock {\em J. Mach. Learn. Res.\/}~{\em 9}, 485–516.

\bibitem[\protect\citeauthoryear{Barabási and Oltvai}{Barabási and
  Oltvai}{2004}]{NetworkBiology}
Barabási, A.-L. and Z.~N. Oltvai (2004).
\newblock Network biology: understanding the cell's functional organization.
\newblock {\em Nature Reviews Genetics\/}~{\em 5\/}(2), 101--113.

\bibitem[\protect\citeauthoryear{Canzler, Schor, Busch, Schubert,
  Rolle-Kampczyk, Seitz, Kamp, von Bergen, Buesen, and Hackermüller}{Canzler
  et~al.}{2020}]{ChallengesMO}
Canzler, S., J.~Schor, W.~Busch, K.~Schubert, U.~E. Rolle-Kampczyk, H.~Seitz,
  H.~Kamp, M.~von Bergen, R.~Buesen, and J.~Hackermüller (2020).
\newblock Prospects and challenges of multi-omics data integration in
  toxicology.
\newblock {\em Archives of Toxicology\/}~{\em 94\/}(2), 371--388.

\bibitem[\protect\citeauthoryear{Chadeau-Hyam, Campanella, Jombart, Bottolo,
  Portengen, Vineis, Liquet, and Vermeulen}{Chadeau-Hyam
  et~al.}{2013}]{Chadeau2013}
Chadeau-Hyam, M., G.~Campanella, T.~Jombart, L.~Bottolo, L.~Portengen,
  P.~Vineis, B.~Liquet, and R.~C. Vermeulen (2013).
\newblock Deciphering the complex: Methodological overview of statistical
  models to derive omics-based biomarkers.
\newblock {\em Environmental and Molecular Mutagenesis\/}~{\em 54\/}(7),
  542--557.

\bibitem[\protect\citeauthoryear{Charbonnier, Chiquet, and
  Ambroise}{Charbonnier et~al.}{2010}]{TimeCourse}
Charbonnier, C., J.~Chiquet, and C.~Ambroise (2010).
\newblock Weighted-lasso for structured network inference from time course
  data.
\newblock {\em Statistical Applications in Genetics and Molecular
  Biology\/}~{\em 9\/}(1).

\bibitem[\protect\citeauthoryear{Chiquet, Mariadassou, and Robin}{Chiquet
  et~al.}{2018}]{PLNnetwork}
Chiquet, J., M.~Mariadassou, and S.~Robin (2018).
\newblock Variational inference for sparse network reconstruction from count
  data.

\bibitem[\protect\citeauthoryear{Erd\"os and R\'enyi}{Erd\"os and
  R\'enyi}{1959}]{ErdosRenyi}
Erd\"os, P. and A.~R\'enyi (1959).
\newblock On random graphs i.
\newblock {\em Publicationes Mathematicae Debrecen\/}~{\em 6}, 290--297.

\bibitem[\protect\citeauthoryear{Foygel and Drton}{Foygel and
  Drton}{2010}]{EBIC}
Foygel, R. and M.~Drton (2010).
\newblock Extended bayesian information criteria for gaussian graphical models.
\newblock In J.~D. Lafferty, C.~K.~I. Williams, J.~Shawe-Taylor, R.~S. Zemel,
  and A.~Culotta (Eds.), {\em Advances in Neural Information Processing Systems
  23}, pp.\  604--612. Curran Associates, Inc.

\bibitem[\protect\citeauthoryear{Friedman, Hastie, and Tibshirani}{Friedman
  et~al.}{2007}]{graphicallassoTibshirani}
Friedman, J., T.~Hastie, and R.~Tibshirani (2007, 12).
\newblock Sparse inverse covariance estimation with the graphical lasso.
\newblock {\em Biostatistics\/}~{\em 9\/}(3), 432--441.

\bibitem[\protect\citeauthoryear{Friedman, Hastie, and Tibshirani}{Friedman
  et~al.}{2010}]{CVlasso}
Friedman, J., T.~Hastie, and R.~Tibshirani (2010).
\newblock Regularization paths for generalized linear models via coordinate
  descent.
\newblock {\em Journal of Statistical Software, Articles\/}~{\em 33\/}(1),
  1--22.

\bibitem[\protect\citeauthoryear{Friedman, Hastie, and Tibshirani}{Friedman
  et~al.}{2018}]{glassopackage}
Friedman, J., T.~Hastie, and R.~Tibshirani (2018).
\newblock {\em glasso: Graphical Lasso: Estimation of Gaussian Graphical
  Models}.
\newblock R package version 1.10.

\bibitem[\protect\citeauthoryear{Giraud}{Giraud}{2008}]{GGMselect1}
Giraud, C. (2008).
\newblock Estimation of gaussian graphs by model selection.
\newblock {\em Electron. J. Statist.\/}~{\em 2}, 542--563.

\bibitem[\protect\citeauthoryear{Guida, Sandanger, Castagné, Campanella,
  Polidoro, Palli, Krogh, Tumino, Sacerdote, Panico, Severi, Kyrtopoulos,
  Georgiadis, Vermeulen, Lund, Vineis, and Chadeau-Hyam}{Guida
  et~al.}{2015}]{Florence}
Guida, F., T.~M. Sandanger, R.~Castagné, G.~Campanella, S.~Polidoro, D.~Palli,
  V.~Krogh, R.~Tumino, C.~Sacerdote, S.~Panico, G.~Severi, S.~A. Kyrtopoulos,
  P.~Georgiadis, R.~C. Vermeulen, E.~Lund, P.~Vineis, and M.~Chadeau-Hyam
  (2015, 01).
\newblock {Dynamics of smoking-induced genome-wide methylation changes with
  time since smoking cessation}.
\newblock {\em Human Molecular Genetics\/}~{\em 24\/}(8), 2349--2359.

\bibitem[\protect\citeauthoryear{Huan, Joehanes, Schurmann, Schramm, Pilling,
  Peters, Mägi, DeMeo, O'Connor, Ferrucci, Teumer, Homuth, Biffar, Völker,
  Herder, Waldenberger, Peters, Zeilinger, Metspalu, Hofman, Uitterlinden,
  Hernandez, Singleton, Bandinelli, Munson, Lin, Benjamin, Esko, Grabe,
  Prokisch, van Meurs, Melzer, and Levy}{Huan et~al.}{2016}]{HuanSmoking}
Huan, T., R.~Joehanes, C.~Schurmann, K.~Schramm, L.~C. Pilling, M.~J. Peters,
  R.~Mägi, D.~DeMeo, G.~T. O'Connor, L.~Ferrucci, A.~Teumer, G.~Homuth,
  R.~Biffar, U.~Völker, C.~Herder, M.~Waldenberger, A.~Peters, S.~Zeilinger,
  A.~Metspalu, A.~Hofman, A.~G. Uitterlinden, D.~G. Hernandez, A.~B. Singleton,
  S.~Bandinelli, P.~J. Munson, H.~Lin, E.~J. Benjamin, T.~Esko, H.~J. Grabe,
  H.~Prokisch, J.~B. van Meurs, D.~Melzer, and D.~Levy (2016, 08).
\newblock {A whole-blood transcriptome meta-analysis identifies gene expression
  signatures of cigarette smoking}.
\newblock {\em Human Molecular Genetics\/}~{\em 25\/}(21), 4611--4623.

\bibitem[\protect\citeauthoryear{Jassal, Matthews, Viteri, Gong, Lorente,
  Fabregat, Sidiropoulos, Cook, Gillespie, Haw, Loney, May, Milacic, Rothfels,
  Sevilla, Shamovsky, Shorser, Varusai, Weiser, Wu, Stein, Hermjakob, and
  D'Eustachio}{Jassal et~al.}{2020}]{Reactome}
Jassal, B., L.~Matthews, G.~Viteri, C.~Gong, P.~Lorente, A.~Fabregat,
  K.~Sidiropoulos, J.~Cook, M.~Gillespie, R.~Haw, F.~Loney, B.~May, M.~Milacic,
  K.~Rothfels, C.~Sevilla, V.~Shamovsky, S.~Shorser, T.~Varusai, J.~Weiser,
  G.~Wu, L.~Stein, H.~Hermjakob, and P.~D'Eustachio (2020).
\newblock The reactome pathway knowledgebase.
\newblock {\em Nucleic Acids Res\/}~{\em 48\/}(D1), D498--d503.

\bibitem[\protect\citeauthoryear{Joehanes, Just, Marioni, Pilling, Reynolds,
  Mandaviya, Guan, Xu, Elks, Aslibekyan, Moreno-Macias, Smith, Brody, Dhingra,
  Yousefi, Pankow, Kunze, Shah, McRae, Lohman, Sha, Absher, Ferrucci, Zhao,
  Demerath, Bressler, Grove, Huan, Liu, Mendelson, Yao, Kiel, Peters,
  Wang-Sattler, Visscher, Wray, Starr, Ding, Rodriguez, Wareham, Irvin, Zhi,
  Barrdahl, Vineis, Ambatipudi, Uitterlinden, Hofman, Schwartz, Colicino, Hou,
  Vokonas, Hernandez, Singleton, Bandinelli, Turner, Ware, Smith, Klengel,
  Binder, Psaty, Taylor, Gharib, Swenson, Liang, DeMeo, O’Connor, Herceg,
  Ressler, Conneely, Sotoodehnia, Kardia, Melzer, Baccarelli, van Meurs,
  Romieu, Arnett, Ong, Liu, Waldenberger, Deary, Fornage, Levy, and
  London}{Joehanes et~al.}{2016}]{LondonSmoking}
Joehanes, R., A.~C. Just, R.~E. Marioni, L.~C. Pilling, L.~M. Reynolds, P.~R.
  Mandaviya, W.~Guan, T.~Xu, C.~E. Elks, S.~Aslibekyan, H.~Moreno-Macias, J.~A.
  Smith, J.~A. Brody, R.~Dhingra, P.~Yousefi, J.~S. Pankow, S.~Kunze, S.~H.
  Shah, A.~F. McRae, K.~Lohman, J.~Sha, D.~M. Absher, L.~Ferrucci, W.~Zhao,
  E.~W. Demerath, J.~Bressler, M.~L. Grove, T.~Huan, C.~Liu, M.~M. Mendelson,
  C.~Yao, D.~P. Kiel, A.~Peters, R.~Wang-Sattler, P.~M. Visscher, N.~R. Wray,
  J.~M. Starr, J.~Ding, C.~J. Rodriguez, N.~J. Wareham, M.~R. Irvin, D.~Zhi,
  M.~Barrdahl, P.~Vineis, S.~Ambatipudi, A.~G. Uitterlinden, A.~Hofman,
  J.~Schwartz, E.~Colicino, L.~Hou, P.~S. Vokonas, D.~G. Hernandez, A.~B.
  Singleton, S.~Bandinelli, S.~T. Turner, E.~B. Ware, A.~K. Smith, T.~Klengel,
  E.~B. Binder, B.~M. Psaty, K.~D. Taylor, S.~A. Gharib, B.~R. Swenson,
  L.~Liang, D.~L. DeMeo, G.~T. O’Connor, Z.~Herceg, K.~J. Ressler, K.~N.
  Conneely, N.~Sotoodehnia, S.~L.~R. Kardia, D.~Melzer, A.~A. Baccarelli,
  J.~B.~J. van Meurs, I.~Romieu, D.~K. Arnett, K.~K. Ong, Y.~Liu,
  M.~Waldenberger, I.~J. Deary, M.~Fornage, D.~Levy, and S.~J. London (2016).
\newblock Epigenetic signatures of cigarette smoking.
\newblock {\em Circulation: Cardiovascular Genetics\/}~{\em 9\/}(5), 436--447.

\bibitem[\protect\citeauthoryear{Jones}{Jones}{2012}]{MethylationMechanisms}
Jones, P.~A. (2012).
\newblock Functions of dna methylation: islands, start sites, gene bodies and
  beyond.
\newblock {\em Nature Reviews Genetics\/}~{\em 13\/}(7), 484--492.

\bibitem[\protect\citeauthoryear{Langfelder and Horvath}{Langfelder and
  Horvath}{2008}]{WGCNA}
Langfelder, P. and S.~Horvath (2008).
\newblock Wgcna: an r package for weighted correlation network analysis.
\newblock {\em BMC Bioinformatics\/}~{\em 9\/}(1), 559.

\bibitem[\protect\citeauthoryear{Leng, Lin, and Wahba}{Leng
  et~al.}{2006}]{accuracycalib}
Leng, C., Y.~Lin, and G.~Wahba (2006).
\newblock A note on the lasso and related procedures in model selection.
\newblock {\em Statistica Sinica\/}~{\em 16\/}(4), 1273--1284.

\bibitem[\protect\citeauthoryear{Liu, Roeder, and Wasserman}{Liu
  et~al.}{2010}]{StARS}
Liu, H., K.~Roeder, and L.~Wasserman (2010).
\newblock Stability approach to regularization selection (stars) for high
  dimensional graphical models.
\newblock In {\em Proceedings of the 23rd International Conference on Neural
  Information Processing Systems - Volume 2}, NIPS’10, Red Hook, NY, USA,
  pp.\  1432–1440. Curran Associates Inc.

\bibitem[\protect\citeauthoryear{Meinshausen and Bühlmann}{Meinshausen and
  Bühlmann}{2006}]{graphicallassoMB}
Meinshausen, N. and P.~Bühlmann (2006, 06).
\newblock High-dimensional graphs and variable selection with the lasso.
\newblock {\em Ann. Statist.\/}~{\em 34\/}(3), 1436--1462.

\bibitem[\protect\citeauthoryear{Meinshausen and Bühlmann}{Meinshausen and
  Bühlmann}{2010}]{stabilityselectionMB}
Meinshausen, N. and P.~Bühlmann (2010).
\newblock Stability selection.
\newblock {\em Journal of the Royal Statistical Society: Series B (Statistical
  Methodology)\/}~{\em 72\/}(4), 417--473.

\bibitem[\protect\citeauthoryear{Müller, Bonneau, and Kurtz}{Müller
  et~al.}{2016}]{StARSpulsar}
Müller, C.~L., R.~A. Bonneau, and Z.~D. Kurtz (2016, May).
\newblock Generalized stability approach for regularized graphical models.
\newblock {\em pre-print\/}.

\bibitem[\protect\citeauthoryear{National Center~for Chronic~Disease,
  on~Smoking, and Health}{National Center~for Chronic~Disease
  et~al.}{2014}]{CDCsmoking}
National Center~for Chronic~Disease, P., H.~P. U.~O. on~Smoking, and Health
  (2014).
\newblock {\em The Health Consequences of Smoking—50 Years of Progress: A
  Report of the Surgeon General}.
\newblock Atlanta (GA): Centers for Disease Control and Prevention (US).

\bibitem[\protect\citeauthoryear{Niedzwiecki, Walker, Vermeulen, Chadeau-Hyam,
  Jones, and Miller}{Niedzwiecki et~al.}{2019}]{Exposome}
Niedzwiecki, M.~M., D.~I. Walker, R.~Vermeulen, M.~Chadeau-Hyam, D.~P. Jones,
  and G.~W. Miller (2019).
\newblock The exposome: Molecules to populations.
\newblock {\em Annu Rev Pharmacol Toxicol\/}~{\em 59}, 107--127.

\bibitem[\protect\citeauthoryear{Noor, Cherkaoui, and Sauer}{Noor
  et~al.}{2019}]{ReviewMO}
Noor, E., S.~Cherkaoui, and U.~Sauer (2019).
\newblock Biological insights through omics data integration.
\newblock {\em Current Opinion in Systems Biology\/}~{\em 15}, 39 -- 47.
\newblock Gene regulation.

\bibitem[\protect\citeauthoryear{Petrovic, Bodinier, Dagnino, Whitaker, Karimi,
  Campanella, Haugdahl~N{\o}st, Polidoro, Palli, Krogh, Tumino, Sacerdote,
  Panico, Lund, Dugu{\'e}, Giles, Severi, Southey, Vineis, Stringhini, Bochud,
  Sandanger, Vermeulen, Guida, and Chadeau-Hyam}{Petrovic et~al.}{2022}]{Dusan}
Petrovic, D., B.~Bodinier, S.~Dagnino, M.~Whitaker, M.~Karimi, G.~Campanella,
  T.~Haugdahl~N{\o}st, S.~Polidoro, D.~Palli, V.~Krogh, R.~Tumino,
  C.~Sacerdote, S.~Panico, E.~Lund, P.-A. Dugu{\'e}, G.~G. Giles, G.~Severi,
  M.~Southey, P.~Vineis, S.~Stringhini, M.~Bochud, T.~M. Sandanger, R.~C.~H.
  Vermeulen, F.~Guida, and M.~Chadeau-Hyam (2022).
\newblock Epigenetic mechanisms of lung carcinogenesis involve differentially
  methylated cpg sites beyond those associated with smoking.
\newblock {\em European Journal of Epidemiology\/}.

\bibitem[\protect\citeauthoryear{Robinson, Kahraman, Law, Lindsay, Nowicka,
  Weber, and Zhou}{Robinson et~al.}{2014}]{DMR}
Robinson, M.~D., A.~Kahraman, C.~W. Law, H.~Lindsay, M.~Nowicka, L.~M. Weber,
  and X.~Zhou (2014).
\newblock Statistical methods for detecting differentially methylated loci and
  regions.
\newblock {\em Frontiers in Genetics\/}~{\em 5}, 324.

\bibitem[\protect\citeauthoryear{Sandanger, Nøst, Guida, Rylander, Campanella,
  Muller, van Dongen, Boomsma, Johansson, Vineis, Vermeulen, Lund, and
  Chadeau-Hyam}{Sandanger et~al.}{2018}]{NOWACIntegration}
Sandanger, T.~M., T.~H. Nøst, F.~Guida, C.~Rylander, G.~Campanella, D.~C.
  Muller, J.~van Dongen, D.~I. Boomsma, M.~Johansson, P.~Vineis, R.~Vermeulen,
  E.~Lund, and M.~Chadeau-Hyam (2018).
\newblock Dna methylation and associated gene expression in blood prior to lung
  cancer diagnosis in the norwegian women and cancer cohort.
\newblock {\em Sci Rep\/}~{\em 8\/}(1), 16714.

\bibitem[\protect\citeauthoryear{Schwarz}{Schwarz}{1978}]{BIC}
Schwarz, G. (1978, 03).
\newblock Estimating the dimension of a model.
\newblock {\em Ann. Statist.\/}~{\em 6\/}(2), 461--464.

\bibitem[\protect\citeauthoryear{Shah and Samworth}{Shah and
  Samworth}{2013}]{stabilityselectionSS}
Shah, R.~D. and R.~J. Samworth (2013).
\newblock Variable selection with error control: another look at stability
  selection.
\newblock {\em Journal of the Royal Statistical Society: Series B (Statistical
  Methodology)\/}~{\em 75\/}(1), 55--80.

\bibitem[\protect\citeauthoryear{Simon, Friedman, Hastie, and Tibshirani}{Simon
  et~al.}{2011}]{CoxLASSO}
Simon, N., J.~Friedman, T.~Hastie, and R.~Tibshirani (2011).
\newblock Regularization paths for cox's proportional hazards model via
  coordinate descent.
\newblock {\em Journal of Statistical Software, Articles\/}~{\em 39\/}(5),
  1--13.

\bibitem[\protect\citeauthoryear{Sustik~M.A.}{Sustik~M.A.}{2012}]{glassoFast}
Sustik~M.A., C.~B. (2012).
\newblock Glassofast: An efficient glasso implementation.
\newblock {\em UTCS Technical Report TR-12-29:1-3\/}.

\bibitem[\protect\citeauthoryear{Tibshirani}{Tibshirani}{1996}]{lassoTibshirani}
Tibshirani, R. (1996).
\newblock Regression shrinkage and selection via the lasso.
\newblock {\em Journal of the Royal Statistical Society. Series B
  (Methodological)\/}~{\em 58\/}(1), 267--288.

\bibitem[\protect\citeauthoryear{Valcárcel, Würtz, Seich~al Basatena,
  Tukiainen, Kangas, Soininen, Järvelin, Ala-Korpela, Ebbels, and
  de~Iorio}{Valcárcel et~al.}{2011}]{ValcarcelDN}
Valcárcel, B., P.~Würtz, N.-K. Seich~al Basatena, T.~Tukiainen, A.~J. Kangas,
  P.~Soininen, M.-R. Järvelin, M.~Ala-Korpela, T.~M. Ebbels, and M.~de~Iorio
  (2011, 09).
\newblock A differential network approach to exploring differences between
  biological states: An application to prediabetes.
\newblock {\em PLOS ONE\/}~{\em 6\/}(9), 1--9.

\bibitem[\protect\citeauthoryear{Vermeulen, Saberi~Hosnijeh, Bodinier,
  Portengen, Liquet, Garrido-Manriquez, Lokhorst, Bergdahl, Kyrtopoulos,
  Johansson, Georgiadis, Melin, Palli, Krogh, Panico, Sacerdote, Tumino,
  Vineis, Castagné, Chadeau-Hyam, on~behalf of~the EnviroGenoMarkers
  Consortium Consortium~members, Botsivali, Chatziioannou, Valavanis,
  Kleinjans, de~Kok, Keun, Athersuch, Kelly, Lenner, Hallmans, Stephanou,
  Myridakis, Kogevinas, Fazzo, De~Santis, Comba, Bendinelli, Kiviranta,
  Rantakokko, Airaksinen, Ruokojarvi, Gilthorpe, Fleming, Fleming, Tu, Lundh,
  Chien, Chen, Lee, Kate~Hsiao, Kuo, Hung, and Liao}{Vermeulen
  et~al.}{2018}]{sPLSLymphoma}
Vermeulen, R., F.~Saberi~Hosnijeh, B.~Bodinier, L.~Portengen, B.~Liquet,
  J.~Garrido-Manriquez, H.~Lokhorst, I.~A. Bergdahl, S.~A. Kyrtopoulos, A.-S.
  Johansson, P.~Georgiadis, B.~Melin, D.~Palli, V.~Krogh, S.~Panico,
  C.~Sacerdote, R.~Tumino, P.~Vineis, R.~Castagné, M.~Chadeau-Hyam, on~behalf
  of~the EnviroGenoMarkers Consortium Consortium~members, M.~Botsivali,
  A.~Chatziioannou, I.~Valavanis, J.~C. Kleinjans, T.~M. de~Kok, H.~C. Keun,
  T.~J. Athersuch, R.~Kelly, P.~Lenner, G.~Hallmans, E.~G. Stephanou,
  A.~Myridakis, M.~Kogevinas, L.~Fazzo, M.~De~Santis, P.~Comba, B.~Bendinelli,
  H.~Kiviranta, P.~Rantakokko, R.~Airaksinen, P.~Ruokojarvi, M.~Gilthorpe,
  S.~Fleming, T.~Fleming, Y.-K. Tu, T.~Lundh, K.-L. Chien, W.~J. Chen, W.-C.
  Lee, C.~Kate~Hsiao, P.-H. Kuo, H.~Hung, and S.-F. Liao (2018).
\newblock Pre-diagnostic blood immune markers, incidence and progression of
  b-cell lymphoma and multiple myeloma: Univariate and functionally informed
  multivariate analyses.
\newblock {\em International Journal of Cancer\/}~{\em 143\/}(6), 1335--1347.

\bibitem[\protect\citeauthoryear{Witten, Friedman, and Simon}{Witten
  et~al.}{2011}]{glasso}
Witten, D.~M., J.~H. Friedman, and N.~Simon (2011).
\newblock New insights and faster computations for the graphical lasso.
\newblock {\em Journal of Computational and Graphical Statistics\/}~{\em
  20\/}(4), 892--900.

\bibitem[\protect\citeauthoryear{Yu}{Yu}{2013}]{stability}
Yu, B. (2013).
\newblock Stability.
\newblock {\em Bernoulli\/}~{\em 19\/}(4), 1484--1500.

\bibitem[\protect\citeauthoryear{Zhao, Roeder, Lafferty, and Wasserman}{Zhao
  et~al.}{2012}]{huge}
Zhao, Roeder, Lafferty, and Wasserman (2012).
\newblock The huge package for high-dimensional undirected graph estimation in
  r.
\newblock {\em Journal of Machine Learning Research\/}~{\em 13}, 1059--1062.

\bibitem[\protect\citeauthoryear{Zou, Hastie, and Tibshirani}{Zou
  et~al.}{2006}]{sPCA}
Zou, H., T.~Hastie, and R.~Tibshirani (2006).
\newblock Sparse principal component analysis.
\newblock {\em Journal of Computational and Graphical Statistics\/}~{\em
  15\/}(2), 265--286.

\end{thebibliography}


\begin{thebibliography}{}

\bibitem[\protect\citeauthoryear{Akaike}{Akaike}{1998}]{AIC}
Akaike, H. (1998).
\newblock {\em Information Theory and an Extension of the Maximum Likelihood
  Principle}, pp.\  199--213.
\newblock New York, NY: Springer New York.

\bibitem[\protect\citeauthoryear{Chiquet, Smith, Grasseau, Matias, and
  Ambroise}{Chiquet et~al.}{2009}]{simone}
Chiquet, J., A.~Smith, G.~Grasseau, C.~Matias, and C.~Ambroise (2009,
  February).
\newblock {SIMoNe: Statistical Inference for MOdular NEtworks.}
\newblock {\em {Bioinformatics -Oxford-}\/}~{\em 25\/}(3), 417--8.

\bibitem[\protect\citeauthoryear{Foygel and Drton}{Foygel and
  Drton}{2010}]{EBIC}
Foygel, R. and M.~Drton (2010).
\newblock Extended bayesian information criteria for gaussian graphical models.
\newblock In J.~D. Lafferty, C.~K.~I. Williams, J.~Shawe-Taylor, R.~S. Zemel,
  and A.~Culotta (Eds.), {\em Advances in Neural Information Processing Systems
  23}, pp.\  604--612. Curran Associates, Inc.

\bibitem[\protect\citeauthoryear{Schwarz}{Schwarz}{1978}]{BIC}
Schwarz, G. (1978, 03).
\newblock Estimating the dimension of a model.
\newblock {\em Ann. Statist.\/}~{\em 6\/}(2), 461--464.

\end{thebibliography}

\end{document}

\newpage

\section{Supplementary Methods: existing calibration strategies}
\label{sec:supp}

In both LASSO-regularised regression and graphical modelling, the calibration of the hyper-parameter $\lambda$ is critical as it regulates the size of the set of selected features. State-of-the-art approaches for the choice of $\lambda$ are based on M-fold cross-validation minimising some error metric (e.g. Mean Squared Error in Prediction). For graphical models, information theory metrics are commonly used, including the Akaike, Bayesian, and Extended Bayesian Information Criterion \citep{AIC, BIC, EBIC, simone}:

\[
\text{AIC}_{\lambda} =  -2 \ell (\hat{\Omega}_{\lambda}) + 2 |E_{\lambda}|
\]

\[
\text{BIC}_{\lambda} = -2 \ell (\hat{\Omega}_{\lambda}) + \log (n) |E_{\lambda}|
\]

\[
\text{EBIC}_{\lambda} = -2 \ell (\hat{\Omega}_{\lambda}) + \log (n) \left( |E_{\lambda}| + 4 \gamma \log (p) \right)
\]

where $\ell (\hat{\Omega}_{\lambda}) = \frac{n}{2}~ \text{log} ~\text{det} ~(\hat{\Omega}_{\lambda}) - \text{tr}~ ( \hat{\Omega}_{\lambda} S)$ is the penalised likelihood, $|E_{\lambda}|$ is the degrees of freedom (i.e. number of edges in the graph), and $\gamma$ is a hyper-parameter specific to the EBIC.

\bibliographystyle{chicago}

\bibliography{biblio}

\clearpage

\section{Supplementary Figures and Tables}
\label{sec:suppfig}

\begin{figure}[h!]
\includegraphics[width=\linewidth]{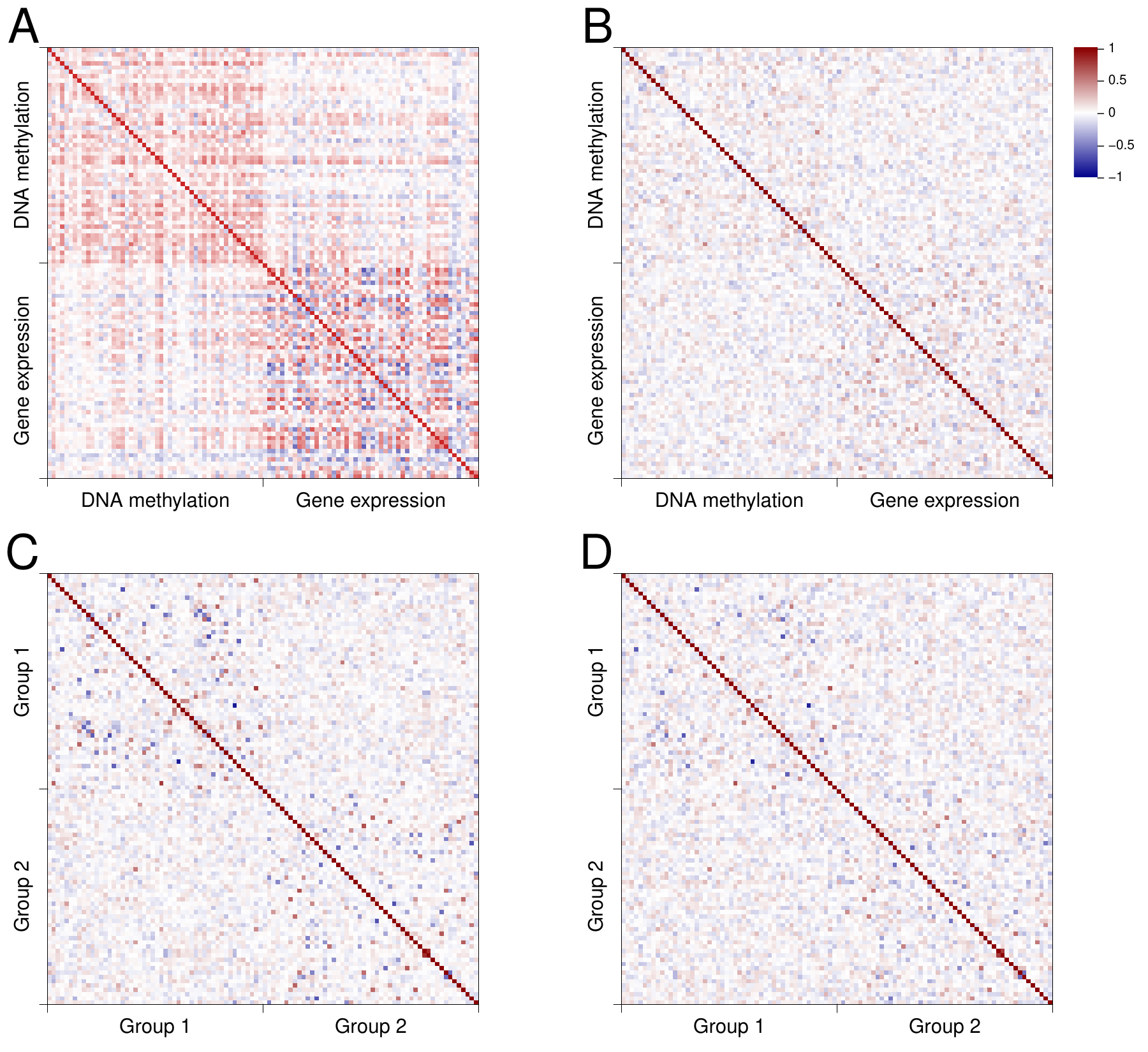} \\
\textbf{Supplementary Figure S1: Simulation of data with a block correlation structure.} The heatmaps show Pearson's correlation (A) and partial correlation (B) matrices estimated on real data from 50 randomly chosen DNA methylation and gene expression markers. The bottom panel shows Pearson's correlation (C) and partial correlation (D) matrices estimated on simulated data with $n=200$ observations and a block structure (50 variables in each group). The simulated conditional independence structure between the $p=100$ variables is that of a random graph ($\nu = 0.04$, $v_b = 0.2$). All partial correlations are estimated without penalisation. 
\end{figure}

\begin{figure}
\begin{center}
\includegraphics[width=\linewidth]{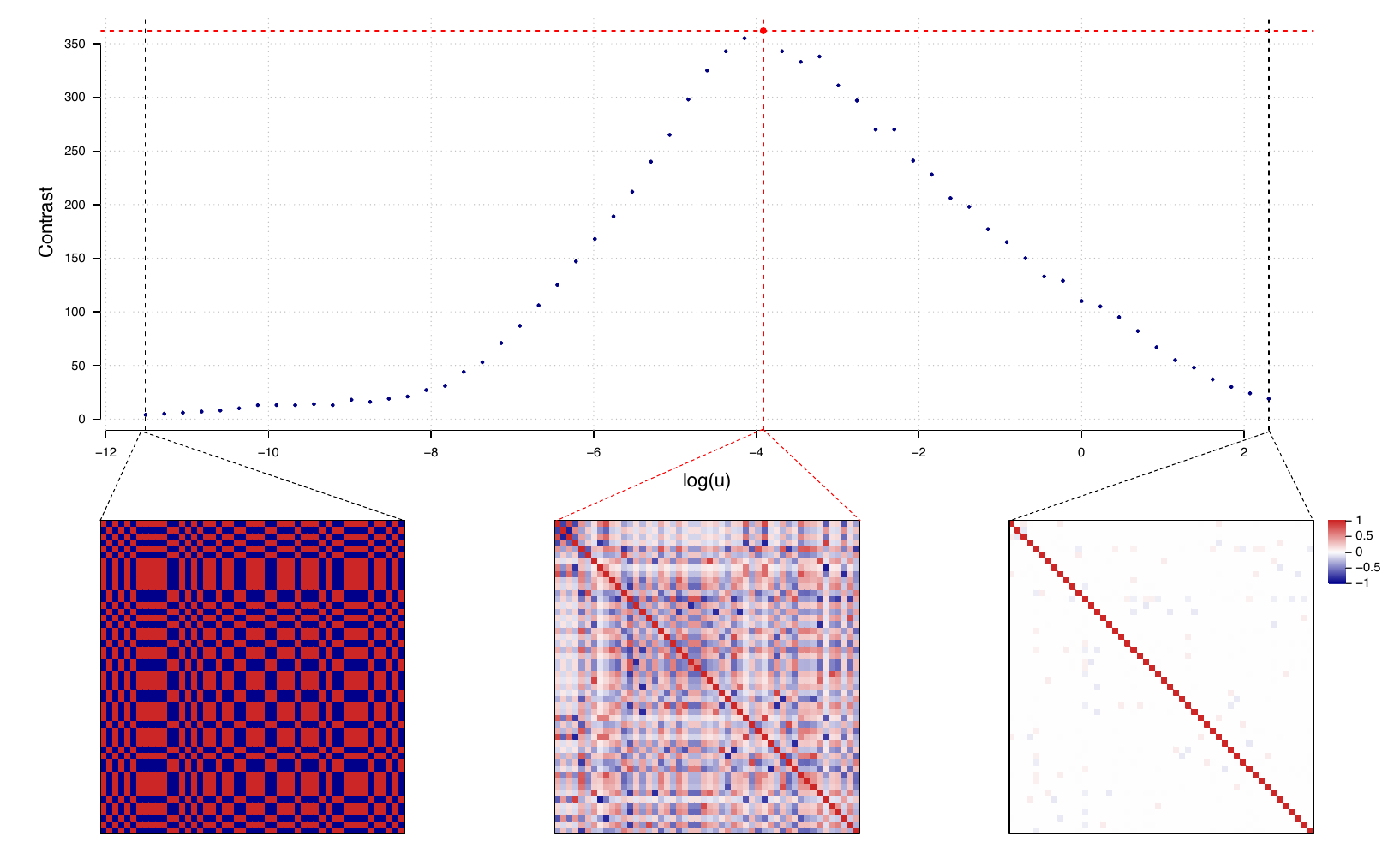} \\
\end{center}
\textbf{Supplementary Figure S2: Choice of the value of parameter $u$ for simulation of the precision matrix.} The contrast of the simulated correlation matrix for a scale-free graphical model with $p = 50$ nodes and $n = 100$ observations is represented as a function of the parameter $u$ on the log-scale. The chosen value for $u$ is the one maximising the contrast (indicated by a red dashed line). The heatmaps of correlation matrices with extreme and calibrated values of the parameter $u$ are showed. 
\end{figure}

\begin{figure}
\begin{center}
\includegraphics[width=\linewidth]{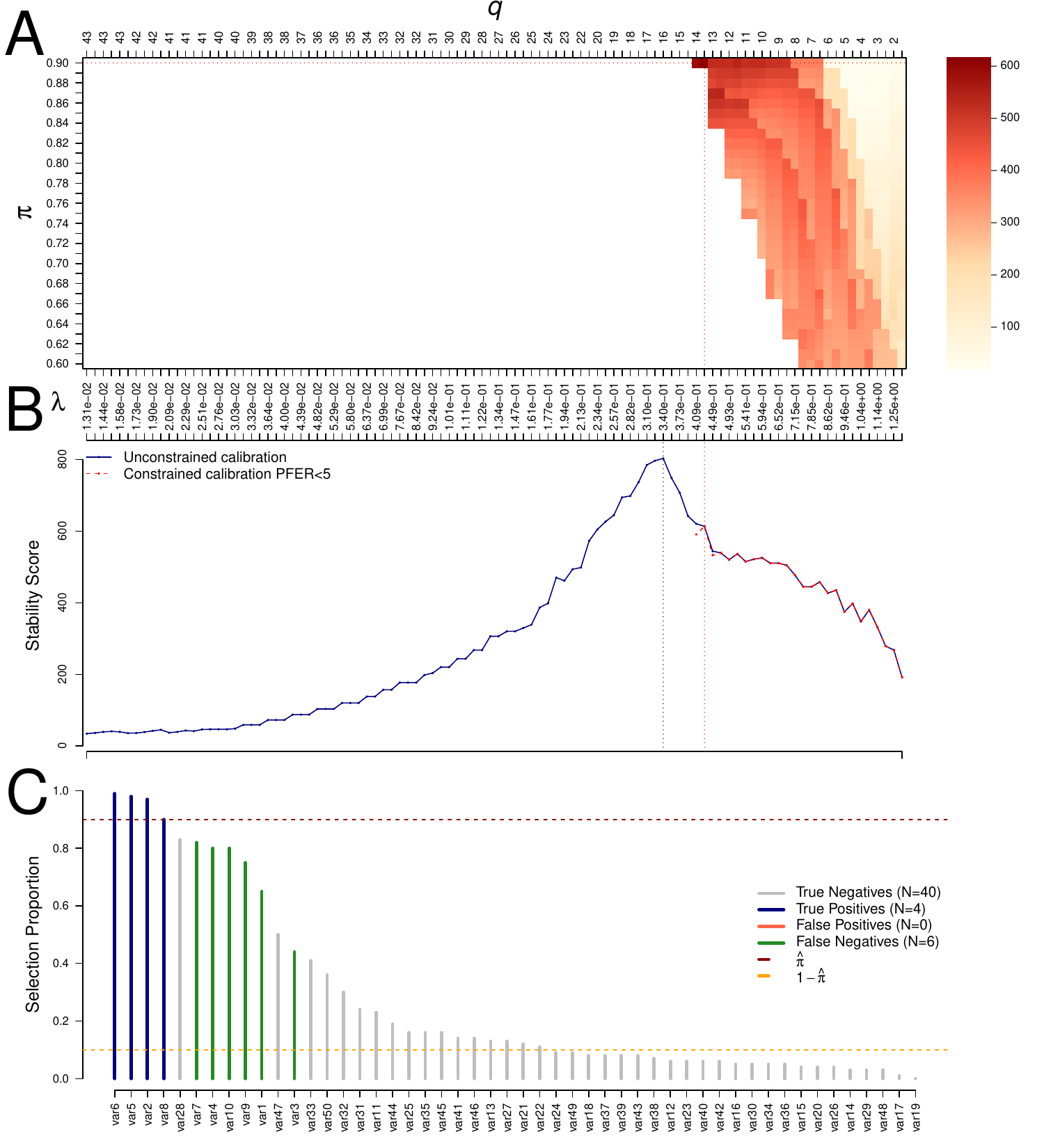} \\
\end{center}
\textbf{Supplementary Figure S3: Visualisation of the PFER constraint in calibration of stability selection models.} The calibration heatmap shows the stability score (colour-coded) as a function of $\lambda$ (or the corresponding average number of selected variables $q$) and $\pi$ (A). The white area (left) represents models for which the PFER computed using the Meinshausen and Bühlmann approach would exceed the threshold ($\text{PFER}_{MB}>5$). The highest stability score obtained for a given penalty parameter $\lambda$ is represented for the unconstrained (blue) and constrained (red dotted line) approaches (B). Ordered selection proportions obtained from constrained calibration are reported (C). Stability selection is applied on simulated data with $n = 100$ observations for $p = 50$ variables, of which $10$ contribute to the definition of the outcome with effect sizes in $\{ [-1, -0.5] \cup [0.5, 1] \}$ and an expected proportion of explained variance of $70\%$. 
\end{figure}

\begin{figure}
\begin{center}
\includegraphics[width=\linewidth]{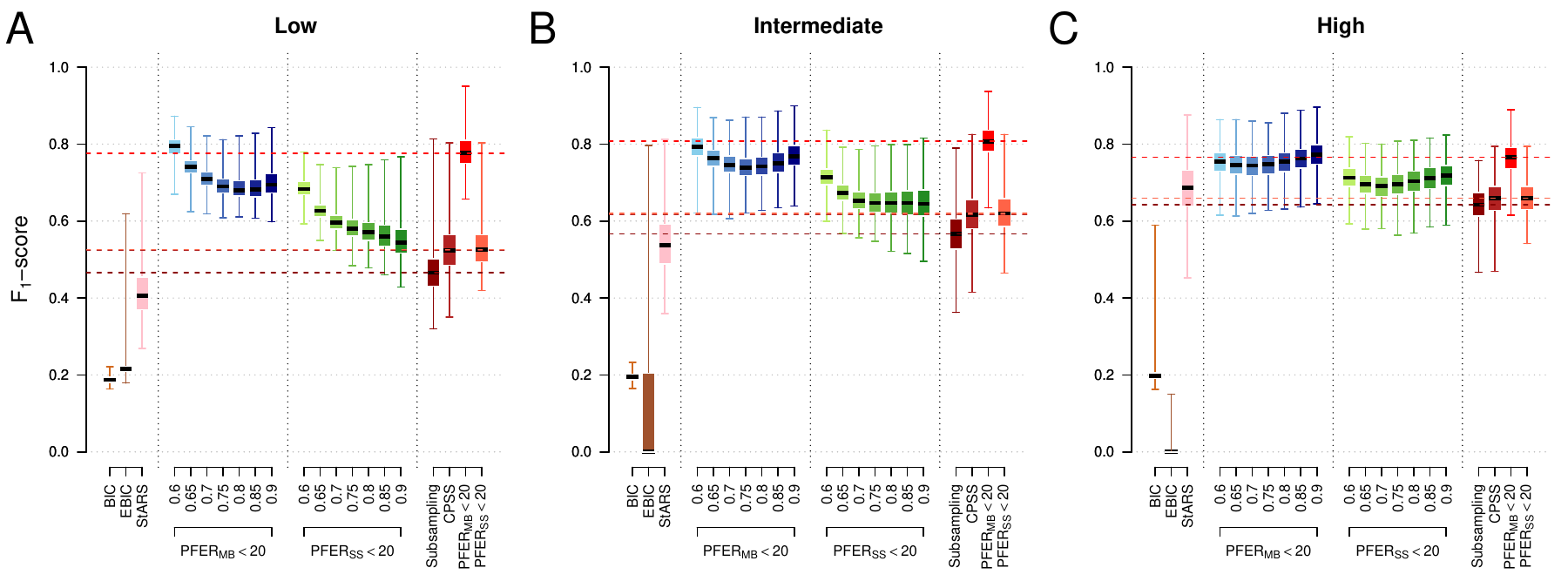} \\
\end{center}
\textbf{Supplementary Figure S4: Selection performances of state-of-the-art approaches and proposed calibrated stability selection graphical LASSO models applied on simulated data with scale-free underlying graph structure.} We show the median, quartiles, minimum and maximum $F_1$-score of graphical LASSO models calibrated using the BIC, EBIC, StARS, and stability selection graphical LASSO models calibrated via error control (MB in blue, SS in green) or using the proposed stability score (red). Models are applied on 1,000 simulated datasets with $p=100$ variables following a multivariate Normal distribution corresponding to a random graph structure ($\nu=0.02$). Performances are estimated in low ($n=2p=200$), intermediate ($n=p=100$), and high ($n=p/2=50$) dimensions.
\end{figure}

\begin{figure}
\begin{center}
\includegraphics[width=0.9\linewidth]{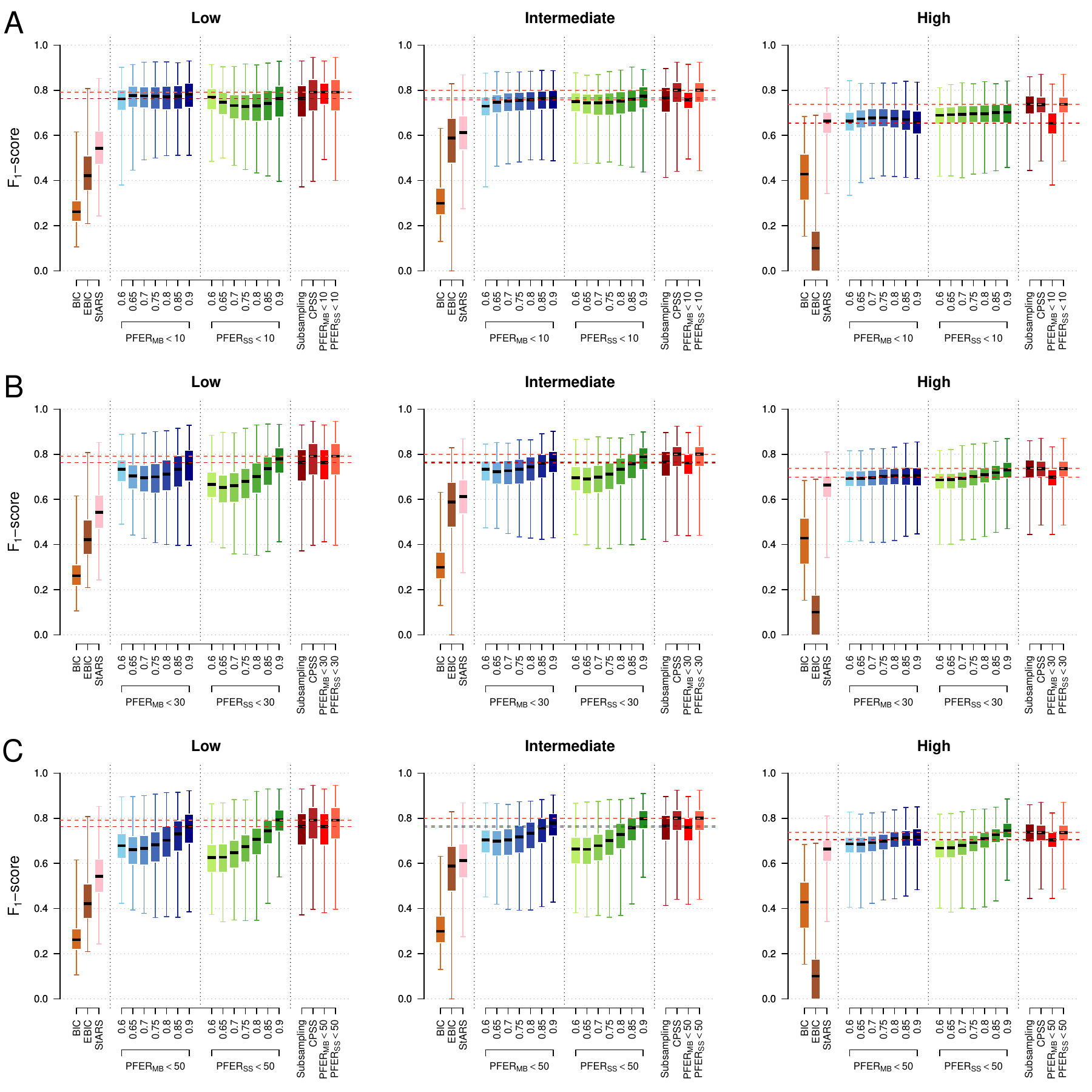} \\
\end{center}
\textbf{Supplementary Figure S5: Selection performances of state-of-the-art approaches and proposed calibrated stability selection graphical LASSO models with different thresholds in PFER.} We show the median, quartiles, minimum and maximum $F_1$-score of graphical LASSO models calibrated using the BIC, EBIC, StARS, and stability selection graphical LASSO models calibrated via error control (MB in blue, SS in green) or using the proposed stability score (red). The threshold in PFER for stability selection models was set to 10 (A), 30 (B) or 50 (C). Models are applied on 1,000 simulated datasets with $p=100$ variables following a multivariate Normal distribution corresponding to a random graph structure ($\nu=0.02$). Performances are estimated in low ($n=2p=200$), intermediate ($n=p=100$), and high ($n=p/2=50$) dimensions. 
\end{figure}
\clearpage

\begin{figure}
\begin{center}
\includegraphics[width=\linewidth]{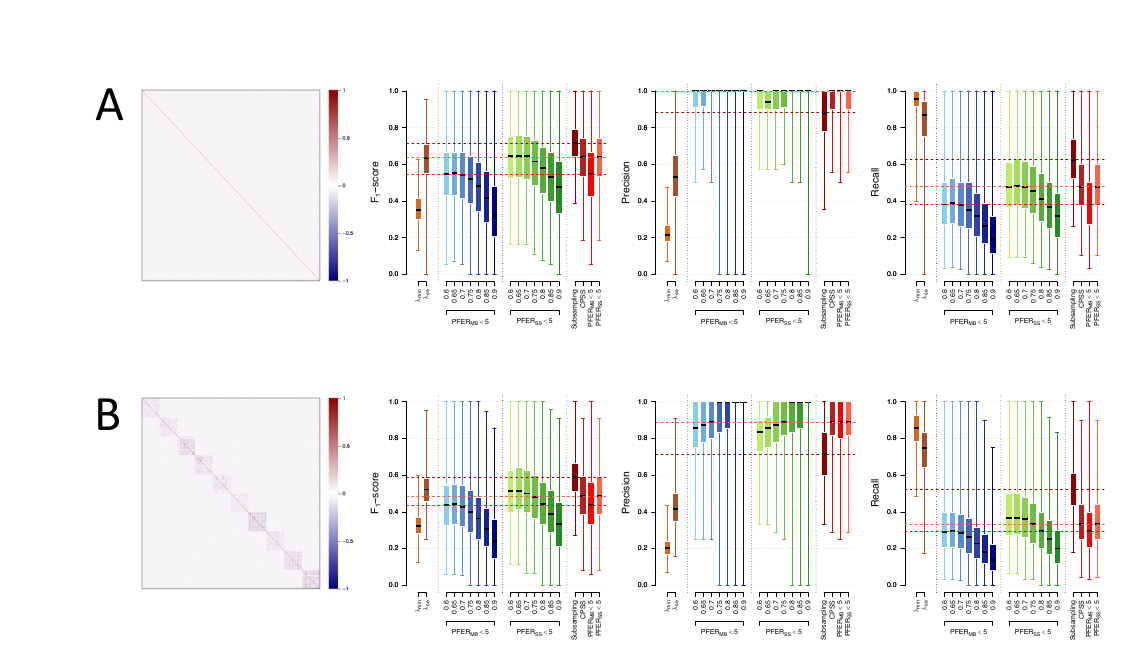} \\
\end{center}
\textbf{Supplementary Figure S6: Selection performances of state-of-the-art approaches and proposed calibrated stability selection LASSO models applied on simulated data.} Models are applied on 1,000 simulated datasets with $n=500$ observations and $p=1,000$ predictor variables, of which an expected proportion $\nu_Y = 0.02$ contributes to the definition of the outcome with effect sizes in $\{-1, 1\}$ and an expected proportion of explained variance of $40\%$. We simulate independent predictors, conditionally on the outcome (A) or blocks of correlated predictors where the conditional independence structure within blocks is that of a random network of density $\nu = 0.02$ (B). For both settings, we represent a heatmap of Pearson's correlations between predictors in a typical simulation. We show the median, quartiles, minimum and maximum $F_1$-score, precision and recall of LASSO models calibrated by 10-fold cross validation minimising the Mean Squared Error of Prediction ($\lambda_{min}$) or one standard error away from the minimum ($\lambda_{se}$), and stability selection LASSO models calibrated via error control (MB in blue, SS in green) or using the proposed stability score (red). 
\end{figure}

\begin{figure}
\begin{center}
\includegraphics[width=\linewidth]{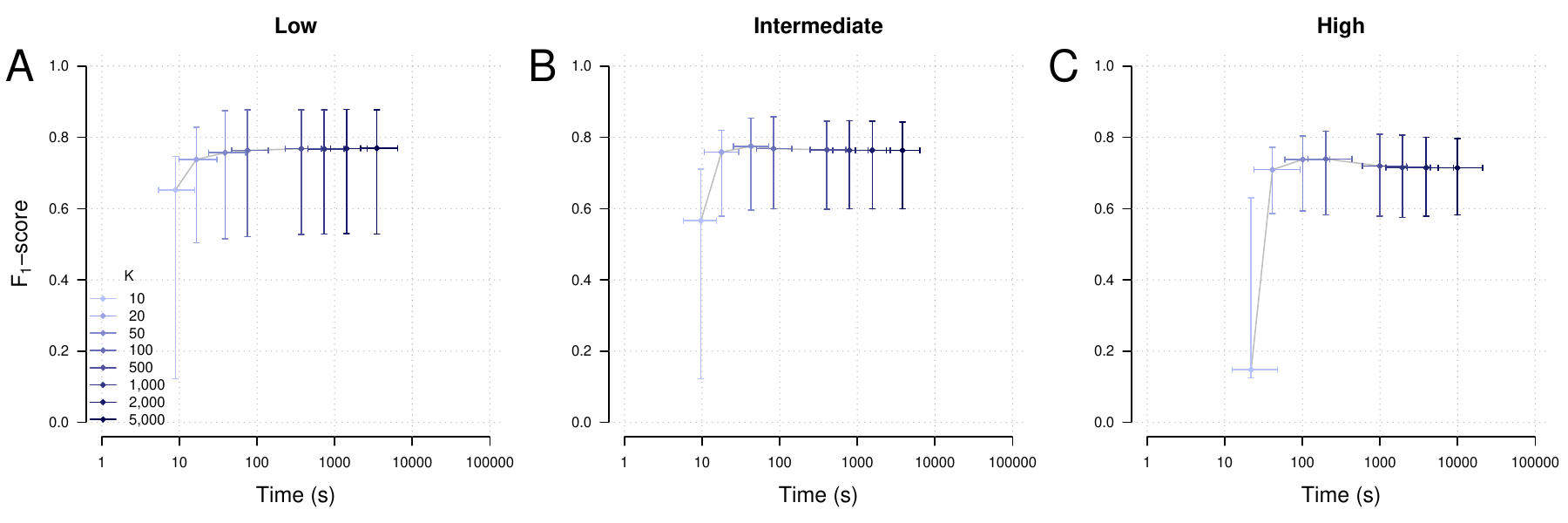} \\
\end{center}
\textbf{Supplementary Figure S7: Effect of the number of subsampling iterations $K$ on the selection performance and computation time.} The median, $5^{th}$ and $95^{th}$ quantile of the $F_1$-score and computation time are reported for graphical LASSO stability selection models calibrated using the unconstrained approach and with different numbers of iterations $K$ (10, 20, 50, 100, 500, 1,000, 2,000, and 5,000). The models are applied on simulated data ($p=100$) with underlying random graph structure ($\nu=0.02$). The computation time in seconds is reported on the log-scale (X-axis). Performances are evaluated in low ($n=2p=200$), intermediate ($n=p=100$), and high ($n=p/2=50$) dimensions.
\end{figure}
\clearpage

\begin{figure}
\begin{center}
\includegraphics[width=0.75\linewidth]{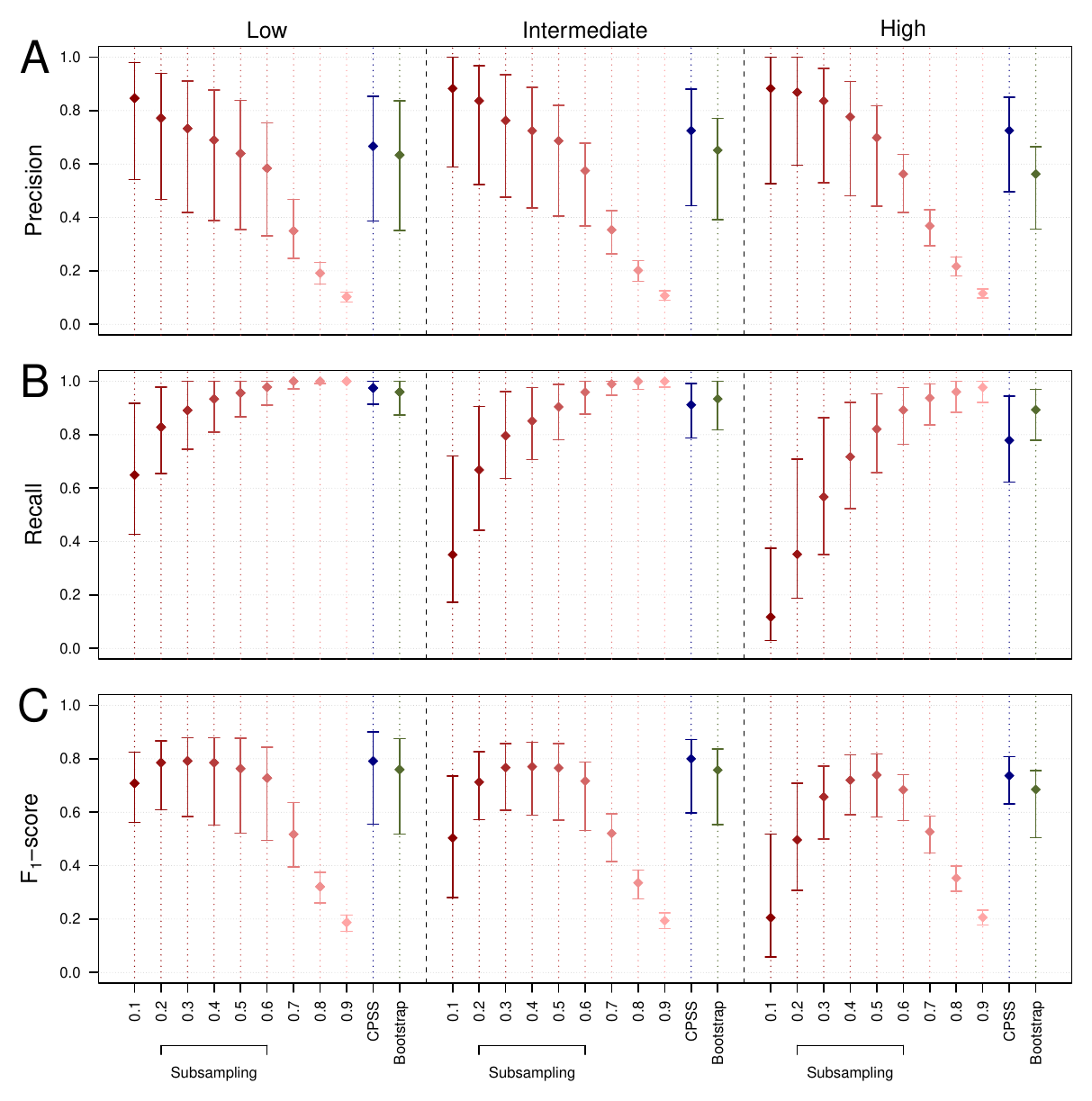} \\
\end{center}
\textbf{Supplementary Figure S8: Effect of the choice of resampling technique on the selection performance.} The median, $5^{th}$ and $95^{th}$ quantile of the $F_1$-score are reported for stability selection models calibrated using the unconstrained approach with different resampling approaches: subsampling with different subsample sizes $\tau$ between 0.1 and 0.9 (red), simultaneous selection in complementary pairs (CPSS, in dark blue) and bootstraping (resampling with replacement, dark green). Performances are evaluated in low ($n=2p=200$), intermediate ($n=p=100$), and high ($n=p/2=50$) dimensions
\end{figure}
\clearpage

\begin{figure}
\begin{center}
\includegraphics[width=0.8\linewidth]{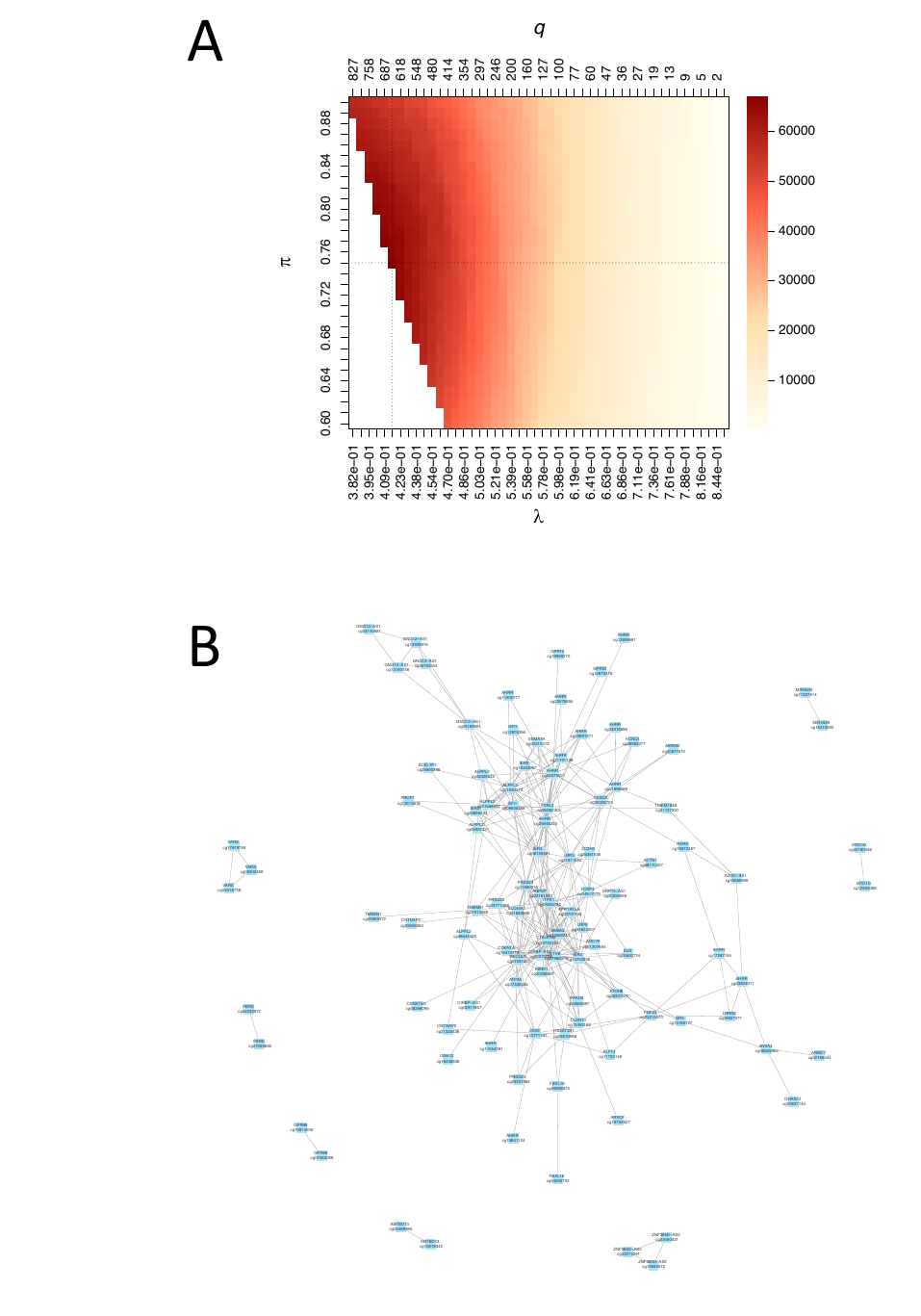} \\
\end{center}
\textbf{Supplementary Figure S9: Single-block graphical model of DNA methylation markers of exposure to tobacco smoking.} Calibration is done by maximising the stability score while ensuring that $\text{PFER}_{MB}<70$ (A). CpG sites with at least one edge are represented in the graph (B). 
\end{figure}
\clearpage

\begin{figure}
\begin{center}
\includegraphics[width=0.8\linewidth]{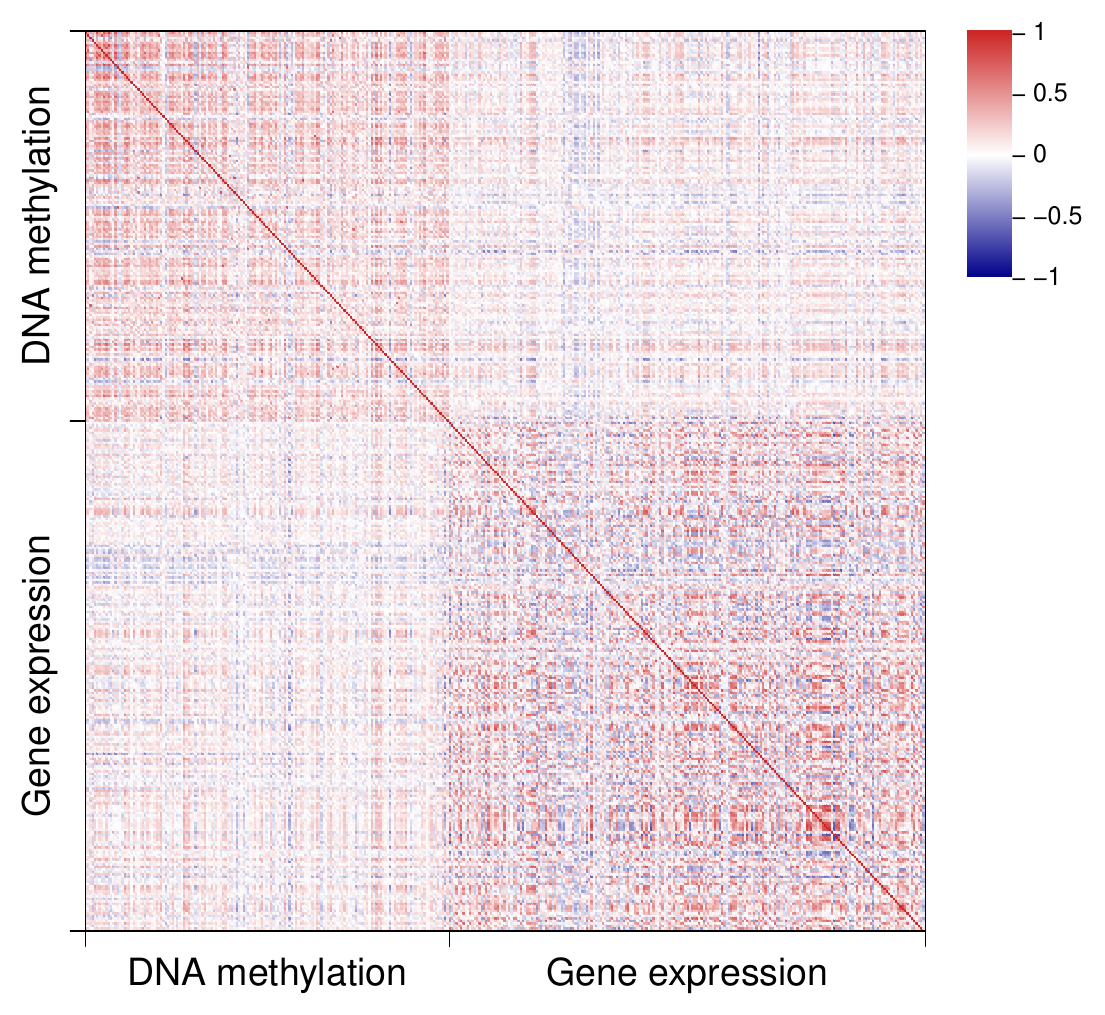} \\
\end{center}
\textbf{Supplementary Figure S10: Heatmap of Pearson's correlations estimated from measured levels of the 159 DNA methylation markers and 208 gene expression markers.}
\end{figure}
\clearpage

\begin{figure}
\begin{center}
\includegraphics[width=0.7\linewidth]{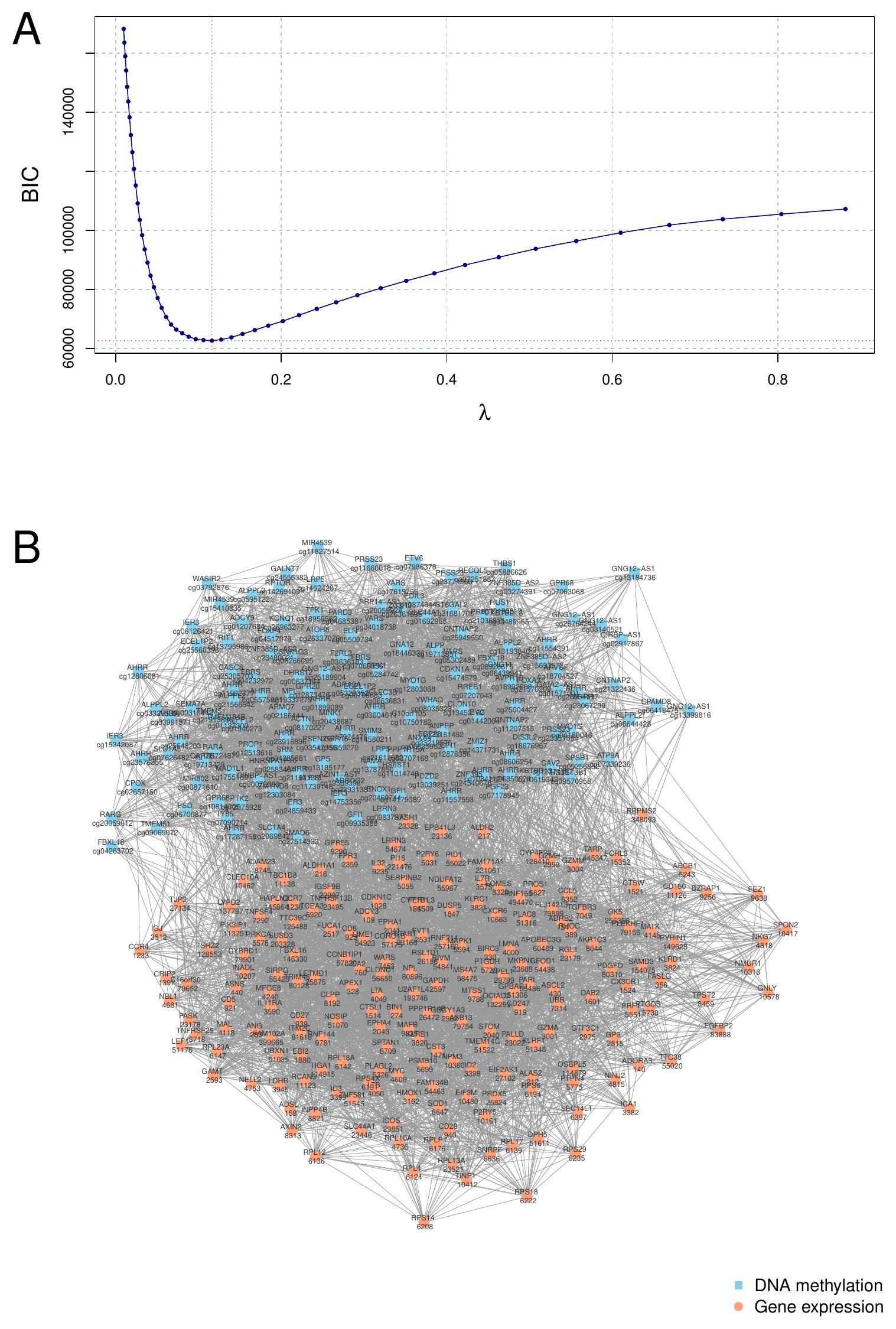} \\
\end{center}
\textbf{Supplementary Figure S11: Graphical LASSO model of smoking-related methylation (blue square) and gene expression (red circle) markers calibrated using the Bayesian Information Criterion (BIC).} The BIC is represented as a function of the penalty parameter $\lambda$ (A). The graphical model generating the smallest BIC is showed (B).
\end{figure}
\clearpage

\begin{figure}
\begin{center}
\includegraphics[width=\linewidth]{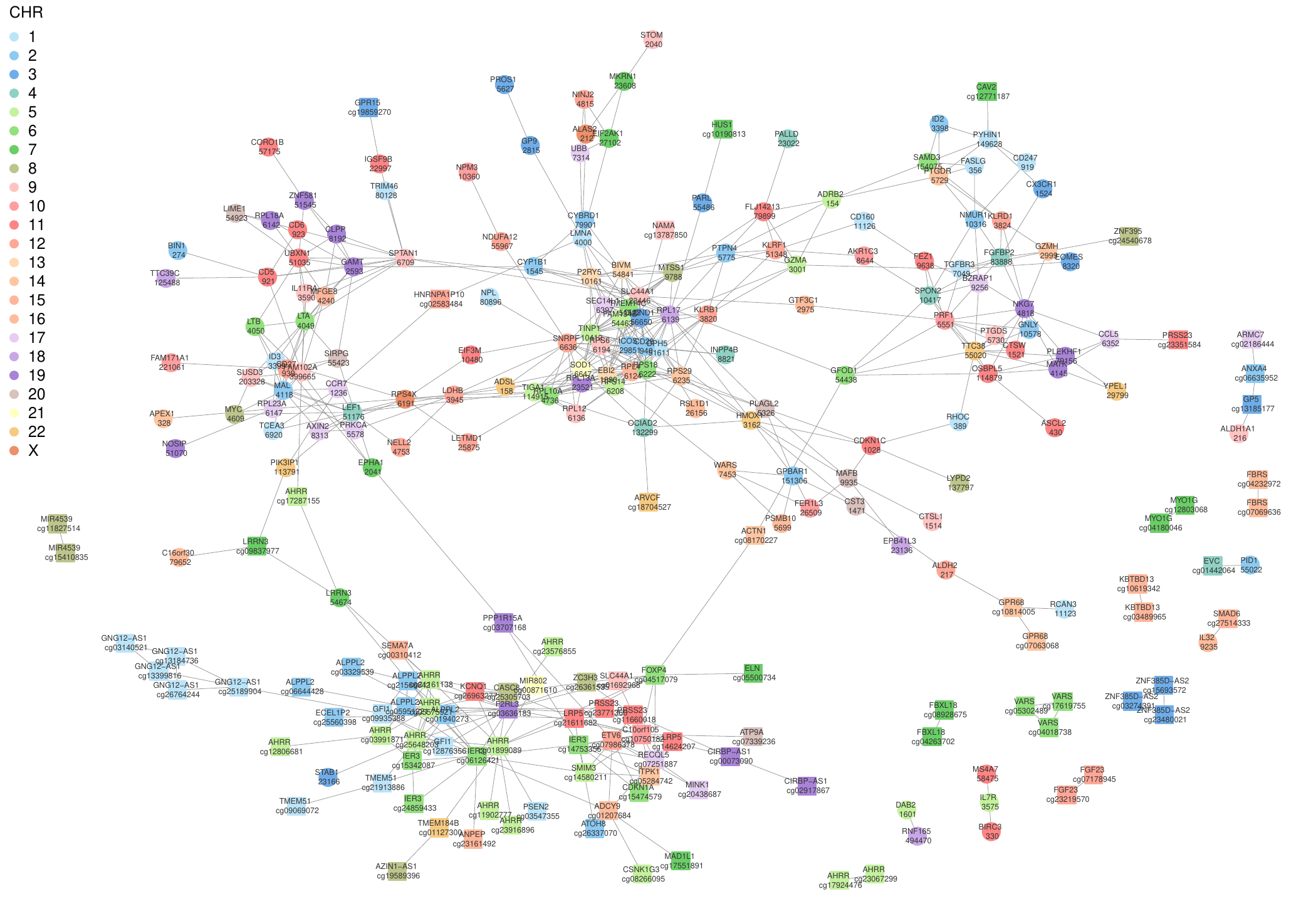} \\
\end{center}
\textbf{Supplementary Figure S12: Multi-OMICs graphical model integrating DNA methylation (square) and gene expression (circle) markers of tobacco smoking with nodes coloured by chromosome.} 
\end{figure}

\clearpage

\noindent
\renewcommand{\arraystretch}{0.8}
{\fontsize{7}{8}\selectfont
\begin{tabular}{cccccccccc} \hline
&&$\pi$&TP&FP&FN&Precision&Recall&$F_1$-score&Time (s)\\ \hline
&AIC&&99 [13]&1959 [448]&0 [0]&0.049 [0.016]&1.000 [0.000]&0.093 [0.030]&1 [0]\\
&BIC&&99 [13]&564 [205]&0 [0]&0.150 [0.061]&1.000 [0.000]&0.260 [0.091]&1 [0]\\
&EBIC&&98 [13]&264 [148]&0 [0]&0.271 [0.125]&1.000 [0.000]&0.426 [0.152]&1 [0]\\
&StARS&&98 [12]&162 [84]&1 [1]&0.373 [0.143]&0.991 [0.012]&0.543 [0.148]&78 [39]\\ \cline{2-10}
&&0.6&89 [12]&44 [19]&10 [7]&0.668 [0.124]&0.897 [0.072]&0.768 [0.093]&83 [40]\\
&&0.65&92 [12]&58 [25]&6 [5]&0.616 [0.123]&0.936 [0.055]&0.743 [0.096]&84 [38]\\
&&0.7&94 [11]&66 [30]&5 [5]&0.590 [0.136]&0.953 [0.048]&0.729 [0.105]&86 [38]\\
&MB&0.75&95 [11]&68 [36]&4 [4]&0.582 [0.150]&0.962 [0.041]&0.725 [0.116]&87 [38]\\
&&0.8&95 [12]&66 [39]&3 [4]&0.590 [0.161]&0.968 [0.038]&0.731 [0.124]&87 [38]\\
&&0.85&95 [10]&62 [40]&3 [5]&0.605 [0.171]&0.969 [0.044]&0.746 [0.127]&88 [38]\\
L&&0.9&95 [10]&56 [41]&3 [5]&0.633 [0.187]&0.968 [0.044]&0.763 [0.131]&88 [37]\\ \cline{2-10}
&&0.6&94 [12]&74 [28]&4 [5]&0.562 [0.123]&0.958 [0.045]&0.708 [0.102]&87 [36]\\
&&0.65&96 [12]&88 [39]&2 [3]&0.522 [0.134]&0.978 [0.031]&0.680 [0.114]&86 [35]\\
&&0.7&97 [11]&90 [47]&2 [3]&0.519 [0.150]&0.981 [0.032]&0.678 [0.129]&86 [34]\\
&SS&0.75&97 [12]&87 [50]&2 [3]&0.531 [0.160]&0.983 [0.029]&0.688 [0.135]&85 [34]\\
&&0.8&97 [12]&80 [50]&1 [3]&0.551 [0.166]&0.989 [0.029]&0.706 [0.136]&84 [32]\\
&&0.85&97 [12]&70 [46]&1 [2]&0.582 [0.170]&0.990 [0.024]&0.733 [0.136]&84 [32]\\
&&0.9&97 [12]&58 [38]&1 [2]&0.631 [0.173]&0.990 [0.021]&0.770 [0.124]&83 [32]\\ \cline{2-10}
&Subsampling&0.9&94 [10]&54 [46]&4 [5]&0.640 [0.203]&0.957 [0.052]&0.764 [0.137]&81 [32]\\
&CPSS&0.9&96 [11]&48 [40]&3 [4]&0.669 [0.199]&0.973 [0.039]&0.793 [0.138]&83 [33]\\
&MB&0.9&93 [10]&52 [39]&5 [5]&0.645 [0.185]&0.949 [0.055]&0.769 [0.126]&81 [32]\\
&SS&0.9&96 [11]&48 [40]&3 [4]&0.669 [0.199]&0.973 [0.039]&0.793 [0.138]&82 [31]\\ \hline
&AIC&&98 [13]&1760 [813]&0 [0]&0.053 [0.028]&1.000 [0.000]&0.101 [0.050]&1 [0]\\
&BIC&&98 [12]&460 [223]&0 [1]&0.176 [0.084]&1.000 [0.010]&0.299 [0.119]&1 [0]\\
&EBIC&&94 [10]&129 [107]&3 [5]&0.426 [0.213]&0.972 [0.050]&0.589 [0.198]&1 [0]\\
&StARS&&94 [10]&115 [75]&3 [5]&0.451 [0.171]&0.966 [0.053]&0.614 [0.151]&82 [42]\\ \cline{2-10}
&&0.6&83 [8]&39 [20]&16 [10]&0.679 [0.128]&0.837 [0.087]&0.748 [0.083]&88 [41]\\
&&0.65&86 [9]&47 [26]&12 [9]&0.648 [0.140]&0.876 [0.079]&0.743 [0.088]&91 [39]\\
&&0.7&87 [8]&50 [30]&11 [9]&0.636 [0.150]&0.893 [0.080]&0.741 [0.092]&95 [40]\\
&MB&0.75&88 [9]&50 [33]&10 [9]&0.640 [0.163]&0.898 [0.084]&0.743 [0.096]&96 [39]\\
&&0.8&88 [9]&47 [34]&10 [9]&0.655 [0.170]&0.901 [0.086]&0.752 [0.097]&95 [38]\\
&&0.85&87 [8]&43 [34]&10 [10]&0.672 [0.176]&0.897 [0.090]&0.761 [0.094]&95 [37]\\
I&&0.9&86 [8]&37 [32]&11 [11]&0.702 [0.176]&0.887 [0.102]&0.773 [0.086]&95 [38]\\ \cline{2-10}
&&0.6&89 [10]&60 [31]&9 [7]&0.600 [0.139]&0.909 [0.072]&0.721 [0.095]&95 [37]\\
&&0.65&91 [10]&66 [39]&7 [7]&0.580 [0.151]&0.931 [0.065]&0.711 [0.106]&95 [35]\\
&&0.7&92 [10]&67 [42]&6 [7]&0.583 [0.161]&0.939 [0.063]&0.716 [0.113]&94 [37]\\
&SS&0.75&92 [10]&63 [43]&6 [7]&0.595 [0.167]&0.941 [0.062]&0.725 [0.114]&94 [38]\\
&&0.8&92 [10]&58 [42]&6 [7]&0.615 [0.174]&0.942 [0.065]&0.738 [0.112]&94 [37]\\
&&0.85&92 [9]&52 [40]&6 [7]&0.642 [0.180]&0.941 [0.069]&0.757 [0.111]&94 [39]\\
&&0.9&91 [9]&44 [35]&6 [7]&0.681 [0.176]&0.941 [0.071]&0.782 [0.098]&94 [37]\\ \cline{2-10}
&Subsampling&0.9&88 [10]&40 [33]&9 [9]&0.687 [0.182]&0.905 [0.086]&0.765 [0.108]&90 [39]\\
&CPSS&0.9&89 [10]&34 [29]&8 [10]&0.725 [0.171]&0.912 [0.088]&0.800 [0.088]&91 [38]\\
&MB&0.89&84 [10]&36 [30]&13 [12]&0.704 [0.166]&0.864 [0.107]&0.763 [0.086]&89 [36]\\
&SS&0.9&89 [10]&34 [29]&8 [10]&0.725 [0.171]&0.912 [0.088]&0.800 [0.088]&89 [35]\\ \hline
&AIC&&97 [12]&3115 [125]&1 [2]&0.030 [0.004]&0.989 [0.022]&0.059 [0.007]&2 [1]\\
&BIC&&93 [9]&242 [240]&4 [8]&0.279 [0.173]&0.961 [0.075]&0.431 [0.203]&2 [1]\\
&EBIC&&5 [10]&0 [1]&92 [12]&0.854 [1.000]&0.054 [0.097]&0.101 [0.175]&2 [1]\\
&StARS&&80 [9]&60 [52]&17 [17]&0.574 [0.185]&0.830 [0.154]&0.664 [0.092]&217 [101]\\ \cline{2-10}
&&0.6&69 [8]&31 [22]&29 [17]&0.694 [0.143]&0.708 [0.140]&0.689 [0.075]&214 [100]\\
&&0.65&72 [9]&34 [25]&26 [17]&0.681 [0.149]&0.733 [0.150]&0.691 [0.072]&211 [105]\\
&&0.7&72 [9]&34 [26]&25 [18]&0.683 [0.150]&0.742 [0.156]&0.695 [0.072]&215 [102]\\
&MB&0.75&72 [10]&32 [27]&26 [19]&0.689 [0.157]&0.737 [0.164]&0.696 [0.075]&214 [103]\\
&&0.8&71 [10]&30 [26]&27 [20]&0.705 [0.161]&0.728 [0.176]&0.697 [0.075]&215 [98]\\
&&0.85&69 [11]&26 [25]&29 [21]&0.726 [0.166]&0.704 [0.187]&0.696 [0.080]&213 [102]\\
H&&0.9&66 [13]&22 [22]&32 [23]&0.753 [0.158]&0.670 [0.196]&0.692 [0.085]&210 [102]\\ \cline{2-10}
&&0.6&76 [8]&43 [28]&21 [16]&0.639 [0.144]&0.780 [0.137]&0.692 [0.072]&206 [100]\\
&&0.65&78 [8]&47 [33]&19 [15]&0.629 [0.153]&0.803 [0.137]&0.692 [0.071]&200 [99]\\
&&0.7&78 [8]&45 [33]&19 [17]&0.636 [0.158]&0.808 [0.145]&0.696 [0.070]&196 [101]\\
&SS&0.75&78 [9]&43 [33]&19 [16]&0.648 [0.163]&0.802 [0.150]&0.699 [0.069]&194 [99]\\
&&0.8&77 [10]&39 [31]&20 [18]&0.667 [0.157]&0.792 [0.160]&0.706 [0.072]&190 [97]\\
&&0.85&76 [10]&34 [29]&21 [19]&0.688 [0.156]&0.782 [0.171]&0.713 [0.071]&190 [99]\\
&&0.9&75 [11]&29 [28]&23 [20]&0.721 [0.156]&0.771 [0.178]&0.720 [0.077]&189 [98]\\ \cline{2-10}
&Subsampling&0.9&80 [10]&35 [23]&17 [13]&0.699 [0.141]&0.822 [0.119]&0.740 [0.079]&189 [95]\\
&CPSS&0.86&76 [9]&29 [18]&22 [15]&0.726 [0.112]&0.779 [0.135]&0.737 [0.067]&185 [89]\\
&MB&0.82&66 [11]&24 [21]&32 [21]&0.733 [0.147]&0.674 [0.175]&0.689 [0.078]&185 [85]\\
&SS&0.86&76 [9]&29 [18]&22 [15]&0.726 [0.112]&0.779 [0.135]&0.737 [0.067]&182 [80]\\ \hline
\end{tabular}}

\textbf{Supplementary Table S1: Median and inter-quartile range of the selection performance metrics and computation times obtained with different graphical models.} Models are applied on 1,000 simulated datasets with $p=100$ variables following a multivariate Normal distribution corresponding to a random graph structure ($\nu=0.02$) in low (L, $n=2p=200$), intermediate (I, $n=p=100$), and high (H, $n=p/2=50$) dimensions.
\clearpage

\noindent
\renewcommand{\arraystretch}{0.8}
\begin{center}
{\fontsize{12}{20}\selectfont 
\begin{tabular}{clcll} \hline
\multicolumn{2}{c}{LASSO} & \multicolumn{3}{c}{Graphical LASSO} \\
p&&p&Cold start&Warm start\\ \hline
1,000&18 [5]&100&69 [33]&51 [22]\\
2,500&35 [9]&250&313 [104]&247 [73]\\
5,000&59 [20]&500&2,759 [1,163]&1,796 [658]\\
7,500&86 [35]&750&14,513 [5,983]&7,402 [3,472]\\
10,000&124 [63]&1,000&99,108 [29,563]&40,240 [10,787]\\ \hline
\end{tabular}}
\end{center}

\textbf{Supplementary Table S2: Median and inter-quartile range of the computation times (in seconds) of stability selection obtained with different numbers of variables $p$.} Models are applied on 1,000 simulated datasets with $n = 500$ observations. For stability selection LASSO models, we use $p = 1,000$, $2,500$, $5,000$, $7,500$ or $10,000$ independent predictors, conditionally on the outcome. For stability selection graphical LASSO models, we use $p=100$, $250$, $500$, $750$ or $1,000$ variables following a multivariate Normal distribution corresponding to a random graph structure ($\nu=0.02$). For graphical models, we report computation times with or without warm start, where models are iteratively fitted over a path from larger to smaller penalty values and the estimate from the previous iteration is a starting point for the gradient descent algorithm (argument "start" in the R package sharp). For LASSO models, we always use warm start as implemented in the R package glmnet.  
\clearpage

\noindent
{\fontsize{7}{12}\selectfont
\begin{tabular}{ccrccccccc} \hline
&&&TP&FP&FN&Precision&Recall&F1-score&Time (s)\\ \hline
&Single&Overall&84 [9]&79 [47]&13 [12]&0.521 [0.143]&0.863 [0.104]&0.643 [0.089]&96 [37]\\ 
&&Within 1&24 [6]&24 [15]&0 [0]&0.500 [0.146]&1.000 [0.000]&0.663 [0.130]&\\ 
&&Between&36 [11]&30 [28]&13 [11]&0.545 [0.191]&0.735 [0.204]&0.604 [0.105]&\\ 
L&&Within 2&24 [6]&24 [15]&0 [0]&0.500 [0.149]&1.000 [0.000]&0.667 [0.127]&\\ \cline{2-10}
&Multi&Overall&93 [10]&73 [34]&6 [6]&0.561 [0.125]&0.941 [0.056]&0.703 [0.087]&269 [94]\\ 
&&Within 1&24 [6]&16 [10]&0 [0]&0.585 [0.132]&1.000 [0.000]&0.737 [0.102]&\\ 
&&Between&44 [8]&38 [25]&5 [5]&0.543 [0.169]&0.893 [0.107]&0.667 [0.110]&\\ 
&&Within 2&24 [6]&16 [10]&0 [0]&0.587 [0.137]&1.000 [0.000]&0.737 [0.106]&\\ \hline
&Single&Overall&77 [10]&62 [40]&22 [15]&0.557 [0.149]&0.782 [0.127]&0.643 [0.066]&107 [45]\\ 
&&Within 1&24 [6]&19 [13]&0 [0]&0.550 [0.170]&1.000 [0.000]&0.704 [0.136]&\\ 
&&Between&29 [11]&22 [22]&21 [14]&0.566 [0.184]&0.582 [0.244]&0.553 [0.107]&\\ 
I&&Within 2&24 [6]&19 [13]&0 [0]&0.554 [0.168]&1.000 [0.000]&0.708 [0.135]&\\ \cline{2-10}
&Multi&Overall&86 [8]&62 [40]&12 [10]&0.583 [0.152]&0.873 [0.080]&0.697 [0.092]&310 [113]\\ 
&&Within 1&24 [7]&11 [9]&0 [1]&0.674 [0.166]&1.000 [0.040]&0.800 [0.113]&\\ 
&&Between&38 [8]&35 [31]&11 [8]&0.526 [0.202]&0.769 [0.146]&0.614 [0.141]&\\ 
&&Within 2&24 [7]&11 [9]&0 [1]&0.676 [0.164]&1.000 [0.037]&0.800 [0.115]&\\ \hline
&Single&Overall&71 [8]&47 [29]&28 [13]&0.597 [0.146]&0.716 [0.109]&0.641 [0.072]&242 [119]\\ 
&&Within 1&24 [7]&14 [11]&0 [1]&0.606 [0.180]&1.000 [0.040]&0.746 [0.131]&\\ 
&&Between&23 [9]&18 [14]&26 [12]&0.559 [0.154]&0.465 [0.198]&0.496 [0.110]&\\ 
H&&Within 2&23 [7]&14 [11]&0 [1]&0.617 [0.176]&1.000 [0.038]&0.750 [0.128]&\\ \cline{2-10}
&Multi&Overall&77 [9]&71 [75]&21 [11]&0.519 [0.214]&0.787 [0.093]&0.627 [0.145]&534 [212]\\ 
&&Within 1&23 [6]&9 [7]&1 [2]&0.719 [0.152]&0.966 [0.082]&0.812 [0.097]&\\ 
&&Between&31 [8]&48 [77]&18 [8]&0.396 [0.306]&0.625 [0.146]&0.480 [0.225]&\\ 
&&Within 2&23 [6]&9 [7]&1 [2]&0.710 [0.150]&0.967 [0.077]&0.812 [0.093]&\\ \hline
\end{tabular}}

\textbf{Supplementary Table S3: Median and inter-quartile range of the selection performance metrics and computation times obtained with single and multi-block stability selection applied on simulated data with a block structure.} For each block, 50 different penalty parameter values are explored. Models are applied on 1,000 simulated datasets with $p=100$ variables following a multivariate Normal distribution corresponding to a random graph ($\nu=0.02$) and with known block structure ($50$ variables per group, using $v_b=0.2$). Performances are evaluated in low (L, $n=2p=200$), intermediate (I, $n=p=100$), and high (H, $n=p/2=50$) dimensions.

\clearpage

\noindent
\renewcommand{\arraystretch}{0.7}
{\fontsize{7}{12}\selectfont
\begin{tabular}{cccccccccc} \hline
&&$\lambda_0$&TP&FP&FN&Precision&Recall&$F_1$-score&Time (s)\\ \hline
S-B&&Overall&85 [9]&78 [44]&14 [12]&0.523 [0.134]&0.859 [0.107]&0.646 [0.081]&70 [30]\\ 
&&Within 1&25 [7]&24 [15]&0 [0]&0.500 [0.152]&1.000 [0.000]&0.667 [0.132]&\\ 
&&Between&36 [11]&29 [27]&14 [11]&0.547 [0.187]&0.725 [0.209]&0.603 [0.099]&\\ 
&&Within 2&24 [7]&23 [14]&0 [0]&0.504 [0.139]&1.000 [0.000]&0.667 [0.123]&\\ \hline
M-P&&Overall&92 [11]&134 [55]&6 [6]&0.408 [0.109]&0.935 [0.060]&0.566 [0.098]&2059 [870]\\ 
&&Within 1&24 [6]&11 [11]&0 [1]&0.679 [0.212]&1.000 [0.050]&0.800 [0.138]&\\ 
&&Between&45 [8]&110 [51]&5 [4]&0.291 [0.094]&0.902 [0.092]&0.438 [0.099]&\\ 
&&Within 2&23 [7]&11 [10]&0 [1]&0.677 [0.188]&1.000 [0.048]&0.794 [0.128]&\\ \hline
M-B&0&Overall&82 [8]&28 [48]&17 [10]&0.750 [0.266]&0.827 [0.094]&0.775 [0.136]&8493 [15464]\\ 
&&Within 1&23 [6]&6 [5]&1 [3]&0.780 [0.148]&0.944 [0.115]&0.844 [0.089]&\\ 
&&Between&36 [7]&14 [43]&13 [9]&0.720 [0.428]&0.737 [0.152]&0.697 [0.220]&\\ 
&&Within 2&23 [7]&6 [5]&1 [3]&0.778 [0.152]&0.944 [0.115]&0.844 [0.091]&\\ \cline{2-10}
&0.001&Overall&83 [8]&33 [65]&15 [11]&0.715 [0.323]&0.845 [0.093]&0.766 [0.184]&2966 [1577]\\ 
&&Within 1&23 [6]&8 [8]&1 [2]&0.750 [0.178]&0.958 [0.094]&0.833 [0.104]&\\ 
&&Between&37 [7]&16 [55]&12 [9]&0.705 [0.473]&0.755 [0.152]&0.689 [0.266]&\\ 
&&Within 2&23 [6]&8 [7]&1 [2]&0.742 [0.169]&0.955 [0.096]&0.828 [0.101]&\\ \cline{2-10}
&0.01&Overall&88 [9]&43 [49]&11 [9]&0.672 [0.230]&0.889 [0.075]&0.760 [0.135]&833 [288]\\ 
&&Within 1&24 [6]&10 [8]&0 [1]&0.699 [0.175]&1.000 [0.045]&0.812 [0.120]&\\ 
&&Between&40 [8]&20 [32]&10 [8]&0.671 [0.315]&0.807 [0.139]&0.714 [0.177]&\\ 
&&Within 2&23 [7]&10 [8]&0 [1]&0.700 [0.165]&1.000 [0.048]&0.811 [0.113]&\\ \cline{2-10}
&0.1&Overall&93 [11]&71 [33]&6 [7]&0.566 [0.123]&0.938 [0.058]&0.705 [0.081]&152 [51]\\ 
&&Within 1&25 [6]&17 [10]&0 [0]&0.591 [0.143]&1.000 [0.000]&0.741 [0.109]&\\ 
&&Between&44 [8]&37 [22]&6 [6]&0.550 [0.158]&0.884 [0.105]&0.671 [0.104]&\\ 
&&Within 2&24 [7]&16 [10]&0 [0]&0.591 [0.133]&1.000 [0.000]&0.742 [0.101]&\\ \cline{2-10}
&0.5&Overall&95 [11]&195 [90]&4 [5]&0.325 [0.107]&0.955 [0.048]&0.486 [0.115]&84 [30]\\ 
&&Within 1&25 [7]&30 [22]&0 [0]&0.451 [0.197]&1.000 [0.000]&0.620 [0.185]&\\ 
&&Between&45 [9]&137 [59]&4 [5]&0.247 [0.082]&0.914 [0.091]&0.389 [0.096]&\\ 
&&Within 2&24 [6]&29 [21]&0 [0]&0.456 [0.188]&1.000 [0.000]&0.624 [0.177]&\\ \cline{2-10}
&1&Overall&95 [11]&225 [99]&4 [5]&0.297 [0.102]&0.954 [0.047]&0.453 [0.114]&75 [25]\\ 
&&Within 1&25 [7]&30 [22]&0 [0]&0.447 [0.191]&1.000 [0.000]&0.618 [0.181]&\\ 
&&Between&45 [9]&165 [69]&4 [5]&0.218 [0.074]&0.914 [0.087]&0.350 [0.094]&\\ 
&&Within 2&24 [6]&29 [22]&0 [0]&0.453 [0.187]&1.000 [0.000]&0.622 [0.176]&\\ \hline
\end{tabular}}

\textbf{Supplementary Table S4: Median and inter-quartile range of the selection performance metrics and computation times obtained with different stability selection models on simulated data with a known block structure.} We compare stability selection not accounting for the block structure (Equation (1), denoted by S-B), using block-specific parameters (Equation (4), M-P) and combining block-specific models calibrated while using different penalties $\lambda_0$ for the other blocks (Equation (5), M-B). For each block, 30 different penalty parameter values are explored. Models are applied on 1,000 simulated datasets with $p=100$ variables following a multivariate Normal distribution corresponding to a random graph ($\nu=0.02$) and with known block structure ($50$ variables per group, using $v_b=0.2$). Performances are evaluated in low dimension ($n=2p=200$). 

\clearpage